\documentstyle[emulateapj,psfig]{article}

\def\freto{f_{\rm ret}^0}
\def\fretinf{f_{\rm ret}^\infty}

\begin{document}

\title{Magnetic Stress at the Marginally Stable Orbit: Altered Disk 
Structure, Radiation, and Black Hole Spin Evolution}

\author{Eric Agol}
\affil{Department of Physics and Astronomy, Johns Hopkins University, 
	   Baltimore, MD 21218, agol@pha.jhu.edu }    
\and
\author{Julian H. Krolik}

\affil{Department of Physics and Astronomy, Johns Hopkins University, 
	   Baltimore, MD 21218, jhk@pha.jhu.edu}

\begin{abstract}

    Magnetic connections to the plunging region can exert stresses
on the inner edge of an accretion disk around a black hole.  We
recompute the relativistic corrections to the thin-disk dynamics equations
when these stresses take the form of a time-steady torque on the inner
edge of the disk. The additional dissipation associated with these stresses
is concentrated relatively close outside the marginally stable orbit,
scaling as $r^{-7/2}$ at large radius.  As a result of these additional
stresses: spin-up of the central black hole is retarded; the maximum
spin-equilibrium accretion efficiency is 36\%, and occurs at $a/M=0.94$;
the disk spectrum is extended toward higher frequencies; line profiles
(such as Fe K$\alpha$) are broadened if the line emissivity scales with
local flux; limb-brightening, especially at the higher frequencies, is
enhanced; and the returning radiation fraction is substantially increased,
up to 58\%.  This last effect creates possible explanations for both 
synchronized continuum fluctuations in AGN, and polarization rises shortward 
of the Lyman edge in quasars.  We show that no matter what additional 
stresses occur, when $a/M < 0.36$, the second law of black hole dynamics 
sets an absolute upper bound on the accretion efficiency.  

\end{abstract}

\keywords{accretion, accretion disks --- black hole physics --- 
galaxies: active --- line profiles ---  polarization --- relativity}

\section{Introduction}

     Early work on black hole accretion disks pointed out the possibility 
that magnetic stresses might exert a torque on the inner parts of the
accretion disk (Page \& Thorne 1974,
Thorne 1974, Ruffini \& Wilson 1975, King \& Lasota 1977).  
However, in virtually every recent account of the dynamics of accretion 
disks around black holes, it has been assumed that there is no stress at 
the disk's inner edge, which should occur very close to the radius of the 
marginally stable orbit, $r_{ms}$.  That this should be so was variously 
argued on the basis that the plunging matter in the region of unstable 
orbits has too little inertia to affect the disk, or rapidly 
becomes causally disconnected from the disk, or that such stresses were
due to relatively weak transport processes that could not compete with
the large gravitational forces pulling matter away from the disk.  
Recently this view has been questioned (Krolik 1999) on the basis that
magnetic fields are the likely agent of torque in accretion disks
(Balbus \& Hawley 1998).   If this is so, and their strength in the
plunging region is what would be expected on the basis of flux-freezing,
they should be strong enough in that zone to both make the Alfv\'en speed
relativistic (postponing the point of causal decoupling) and exert
forces competitive with gravity.  

     If matter inside the marginally stable orbit does, indeed, remain
magnetically connected to the disk, it can exert a sizable torque on
the the portion of the disk containing the field-line footpoints.
Gammie (1999) has shown that, within the confines of a highly-idealized model
of inflow dynamics, this torque can considerably enhance the amount
of energy released in the disk.

      In fact, even if there were no continuing accretion, field lines
attached to the event horizon of a spinning black hole and running through
the disk could exert torques of a very similar character (Blandford 1998,
D.M. Eardley, private communication).  We will call this situation the
``infinite efficiency limit."

      A corollary of torque on the inner edge of the disk is an increase
in the outward angular momentum flux.  In a time-steady state, this
additional angular momentum flux must be conveyed by additional stress.
Additional local dissipation must accompany the additional stress.  It is
the principal object of this paper to compute how this dissipation is
distributed through the disk, and examine the consequences for observable
properties.

      Time-steady torques at $r_{ms}$ are not the only way that energy
may be transmitted from the plunging region to the disk---the torque may
be variable, it may be delivered over a range of radii, and there may
be radial forces exerted that carry no angular momentum.  However,
in this paper, we will restrict our attention to this simplest possible
case.

\section{The Relativistic Correction Factors}

\subsection{Dissipation as a function of radius}

     Novikov \& Thorne (1973) and Page \& Thorne (1974) showed how the surface
brightness and vertically-integrated stress in the fluid frame for a
time-steady, geometrically thin, relativistic accretion
disk could be written as the Newtonian forms multiplied by correction
factors that approach unity at large radius.  In the notation of
Page \& Thorne (1974), conservation of angular momentum is given by
\begin{equation}
{\partial \over \partial r} \left(L^\dagger +
{C^{1/2} \over B\partial\Omega/\partial r} f\right) =  L^\dagger f,
\end{equation}
where $r$ is the Boyer-Lindquist radial coordinate, $L^\dagger$ is the conserved
specific angular momentum of a circular orbit at radius $r$, $f$
is a function of radius defined such that the flux at the disk surface in
the fluid frame $F = \dot M_o f/(4\pi r)$, and $\dot M_o$ is the rest-mass
accretion rate.  As usual, $\Omega$ is the
angular frequency of a circular orbit at radius $r$.  We also follow
Novikov \& Thorne (1973) by defining four auxiliary functions:
\begin{eqnarray}
B(x) = 1 + a_*/x^{3/2} \\
C(x) = 1 - 3/x + 2a_*/x^{3/2} \\
D(x) = 1 - 2/x + a_*^2/x^2 \\
F(x) = 1 - 2a_*/x^{3/2} + a_{*}^2/x^2 ,
\end{eqnarray}
with $x$ the radius in units of $r_g=GM/c^2$ and $a_*$ the dimensionless
black hole spin parameter.  In the usual approach, the
boundary condition on $f$ at the radius $r_{ms}$ of the marginally stable orbit
is $f_{ms} = 0$.  The appropriate boundary condition when
there is non-zero stress at $r_{ms}$ is
\begin{equation}
f_{ms} = {3 \over 2} {\Delta \epsilon \over x_{ms} C_{ms}^{1/2}},
\end{equation}
where $C_{ms}=C(r_{ms})$, and $\Delta \epsilon$ is the additional radiative 
efficiency relative
to the one computed in terms of the binding energy at $r_{ms}$, $\epsilon_0$,
so that $\epsilon = \Delta \epsilon + \epsilon_0$.  This choice of $f_{ms}$ 
ensures that the integrated additional dissipation matches $\Delta\epsilon$,
and corresponds to a stress
\begin{equation}
W^r_\phi(r_{ms}) = {\Delta \epsilon \dot M_o \over 2 \pi r_{ms} \Omega_{ms}}.
\end{equation}
We refer to a disk with $\Delta \epsilon =0$ as a ``Novikov-Thorne disk.''  

    Using this boundary condition, the locally generated surface flux becomes
\begin{equation} \label{flux}
F(x) = {3 \over 8\pi}{GM \dot M_o \over r^3}\left[ {x_{ms}^{3/2} C_{ms}^{1/2}
\Delta\epsilon \over C(x) x^{1/2}} + R^{NT}_R(x)\right]
\end{equation}
where $R^{NT}_R(x)$ is the expression found by Novikov \& Thorne (1973).
The standard relativistic 
correction factor $R^{NT}_R$ goes to zero as $x$ approaches $x_{ms}$ from above,
so that $F$ (when the inner-edge stress is zero) peaks well outside the 
marginally stable orbit.  By contrast, the additional dissipation due to a 
torque on the inner edge is concentrated very close to $r_{ms}$, and is non-zero
at the inner edge.  The degree of concentration can be quantified by 
measuring $r_{1/2}$, the radius within which fifty percent of the radiation 
is emitted: the half-light radius is a factor of a few smaller for torque-driven
flux than for Novikov-Thorne flux.
Figure 1 shows the half-light radius for a Novikov-Thorne disk, $r^0_{1/2}$, 
and an infinite-efficiency disk, $r^\infty_{1/2}$, as a function of $a_*$.

In the limit of infinite efficiency or zero accretion rate,
$\dot M_o\epsilon$ remains finite, so the first term in equation [8] dominates; 
in this case, 
the flux scales as $r^{-7/2}$ at large $r$ rather than as $r^{-3}$ as in the 
standard thin disk. The expression for the surface flux becomes:
\begin{equation}
F^{(\infty)}(x) = {3 \over 2}{c^3 \over r_g \kappa_{T} x^{7/2}}{L \over L_{Edd}}
{x_{ms}^{3/2} C_{ms}^{1/2} \over C(x)}, 
\end{equation}
where $L_{Edd}$ is the Eddington luminosity and $\kappa_{T}$ is the Thomson 
opacity per unit mass. 
\vskip 2mm
\hbox{~}
\centerline{\psfig{file=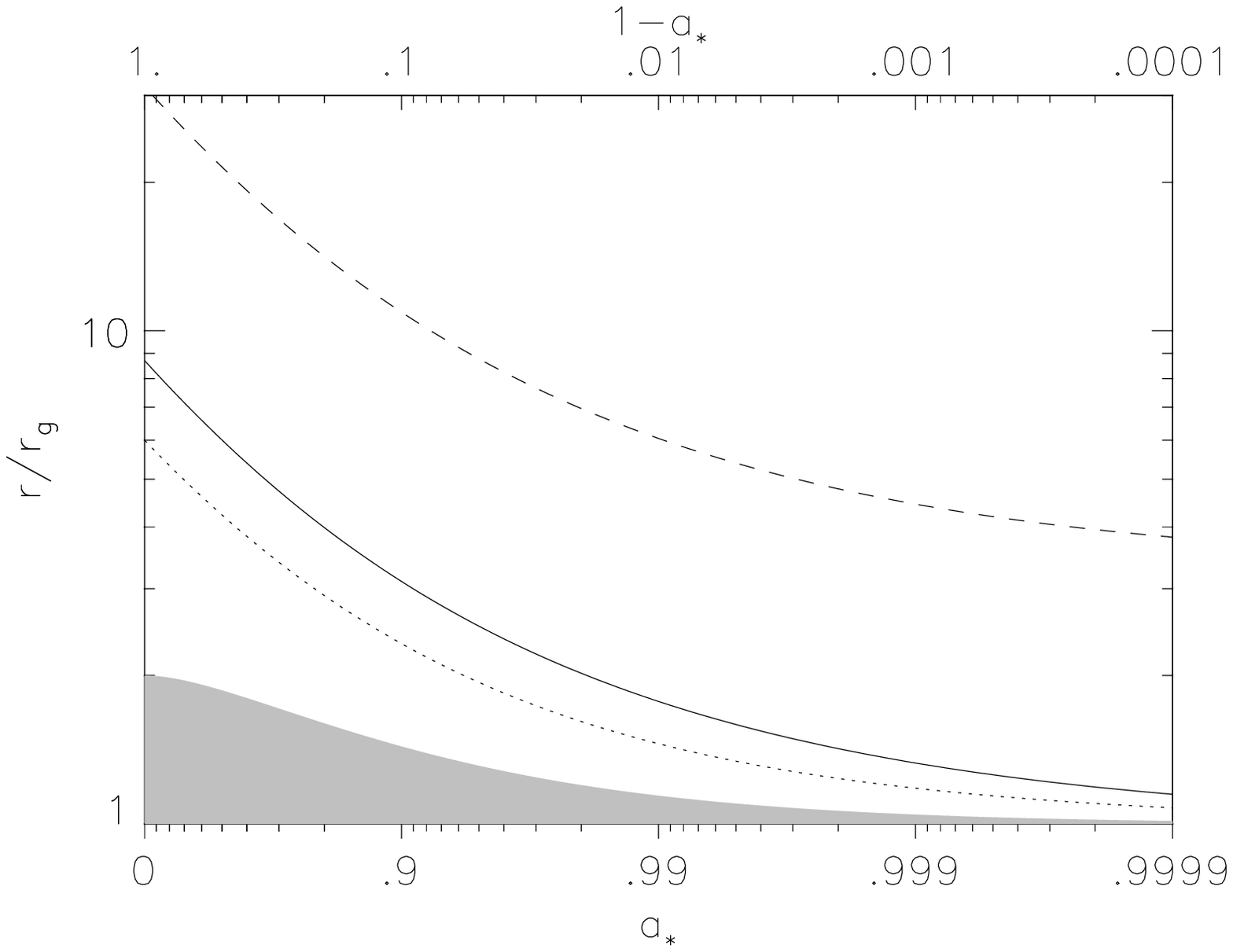,width=3.6in}} 
\noindent{
\scriptsize \addtolength{\baselineskip}{-3pt}
\vskip 1mm
\begin{normalsize}
Fig.~1.\ Half-light radii for the Novikov-Thorne disk (dashed line)
and infinite-efficiency disk (solid line).  Also plotted are $r_{ms}$ (dotted
line) and the region inside the horizon (shaded).
\end{normalsize}
\vskip 3mm
\addtolength{\baselineskip}{3pt}
}

     The angular momentum conservation equation corresponding to equation [8]
is
\begin{equation}
-\int dz T_{r\phi}(z) = {\dot M_o \Omega_K(x) \over 2 \pi} 
\left[ {x_{ms}^{3/2} C_{ms}^{1/2} \Delta\epsilon \over D(x) x^{1/2}} 
+ R^{NT}_T(x)\right],
\end{equation}
where $R^{NT}_T$ is the torque correction factor (Novikov \& Thorne 1973,
Page \& Thorne 1974) in the notation of Krolik (1999).

    To clarify the meaning of the extra dissipation, we will write
down equation [8] pretending that gravity is purely Newtonian:
\begin{equation}
F^{(N)}(r) = {3 \over 8\pi}{GM \dot M_o \over r^3} \left[\Delta \epsilon \sqrt{r_{in} 
\over r} {r_{in}\over r_g}+\left(1-\sqrt{r_{in}\over r}\right) \right],
\end{equation}
where $r_{in}$ is the disk inner edge.
The first term in the bracket is the usual Shakura-Sunyaev (1973)
correction factor, while the second term is derived from the extra torque
at the inner edge.
This equation (actually first derived by Popham \& Narayan 1993) never applies in 
the relativistic case, but can apply, for example, to a disk around a star where 
a torque is exerted by the spinning magnetosphere or through a boundary layer,
or to a thin disk surrounding
a different disk solution, such as an ADAF, where a torque is exerted
by the flow inside the transition point.

\subsection{Returning radiation}

    To find the surface brightness distribution of the disk as seen by distant
observers, it is necessary first to correct the intrinsic surface brightness
due to local dissipation for the additional energy
supplied by photons originally emitted at a different radius, but returned to
the disk by gravity.  In the conventional picture, this is a small correction
(Cunningham 1976).  Here, however, because so much more of the energy is
released deep in the relativistic potential, it can be a much greater effect.

    To compute the additional returning radiation, we followed the method
developed by Cunningham (1976), with a few modifications.  The numerical
method is described in Agol (1997).
We compute the flux transfer function, $T_f$, by following photons emitted from
each radius that return to the accretion disk, assuming the disk surface
is flat and the radiation is isotropic in the fluid frame.  We ignore the 
stress carried
by these photons (i.e., we set $T_s=0$ in Cunningham's parlance).  We also
assume that any radiation that returns to the disk inside $r_{ms}$ is
captured by the black hole---this radiation will be advected or scattered
inwards by the inflowing gas, which has a large inward radial velocity.  
Finally, we (temporarily) assume (as does Cunningham) that the radiation
returning to the disk is absorbed and thermalized before being reemitted;
this assumption is probably not appropriate in practice, but greatly
simplifies computation of the transfer function since in this approximation
$T_f$ is independent of frequency.  We will discuss later how breaking
this assumption may change the spectrum.
    
    In Figure 2a, we plot versus radius the fraction of emitted radiation
which returns to the disk outside $r_{ms}$, which enters the black
hole or returns to the disk inside $r_{ms}$, and which reaches infinity
directly, for the cases $a_*=0.9999$ and $a_*=0$.  The fraction
reaching infinity and returning to the disk are nearly independent of the
black hole spin.  The fraction returning to the disk is greater than 10\%
for $r \lesssim 6r_g$, so when the emitted energy is concentrated inside
this radius (see figure 1), then returning radiation will play an important
role in modifying disk spectra.  For $r \lesssim 1.5 r_g$, less than half
of the radiation reaches infinity directly - most returns to the disk.  For
$a_*=0$, the fraction of radiation which is captured by the black hole or
returns inside $r_{ms}$ increases since $r_{ms}$ is so large; this
fraction never exceeds 8\%.

    The fraction of returning radiation integrated over all radii, 
$f_{\rm ret}$ (as measured at
infinity), is shown (as a function of $a_*$) in Figure 2b (dashed lines)
for two limiting efficiencies: $\epsilon=\epsilon_0$ ($\freto$) and
$\epsilon=\infty$ ($\fretinf$).  
The fraction for any other efficiency can be found by taking a linear
combination of the fraction for these two efficiencies
\begin{equation} \label{fraction}
f_{\rm ret} = (f_{\rm ret}^0 \epsilon_0 + f_{\rm ret}^\infty 
\Delta\epsilon)/\epsilon.
\end{equation}
As can be seen from the figure, $f_{\rm ret}$ is relatively small for
$a_* = 0$, even for $\epsilon=\infty$.  However, $f_{\rm ret}$
grows quickly with increasing $a_*$.  The primary reason for this is that
$r_{ms}$ shrinks with increasing $a_*$, so that relativistic effects on
the photon trajectories become more important.  Trajectory curvature is
especially strong for those photons coming from small radii whose initial
direction would carry them over the black hole.  When $a_*$ and
$\Delta\epsilon$ are comparatively large, up to 58\% of 
the energy due to the extra dissipation ends up striking the disk.
\vskip 2mm
\hbox{~}
\centerline{\psfig{file=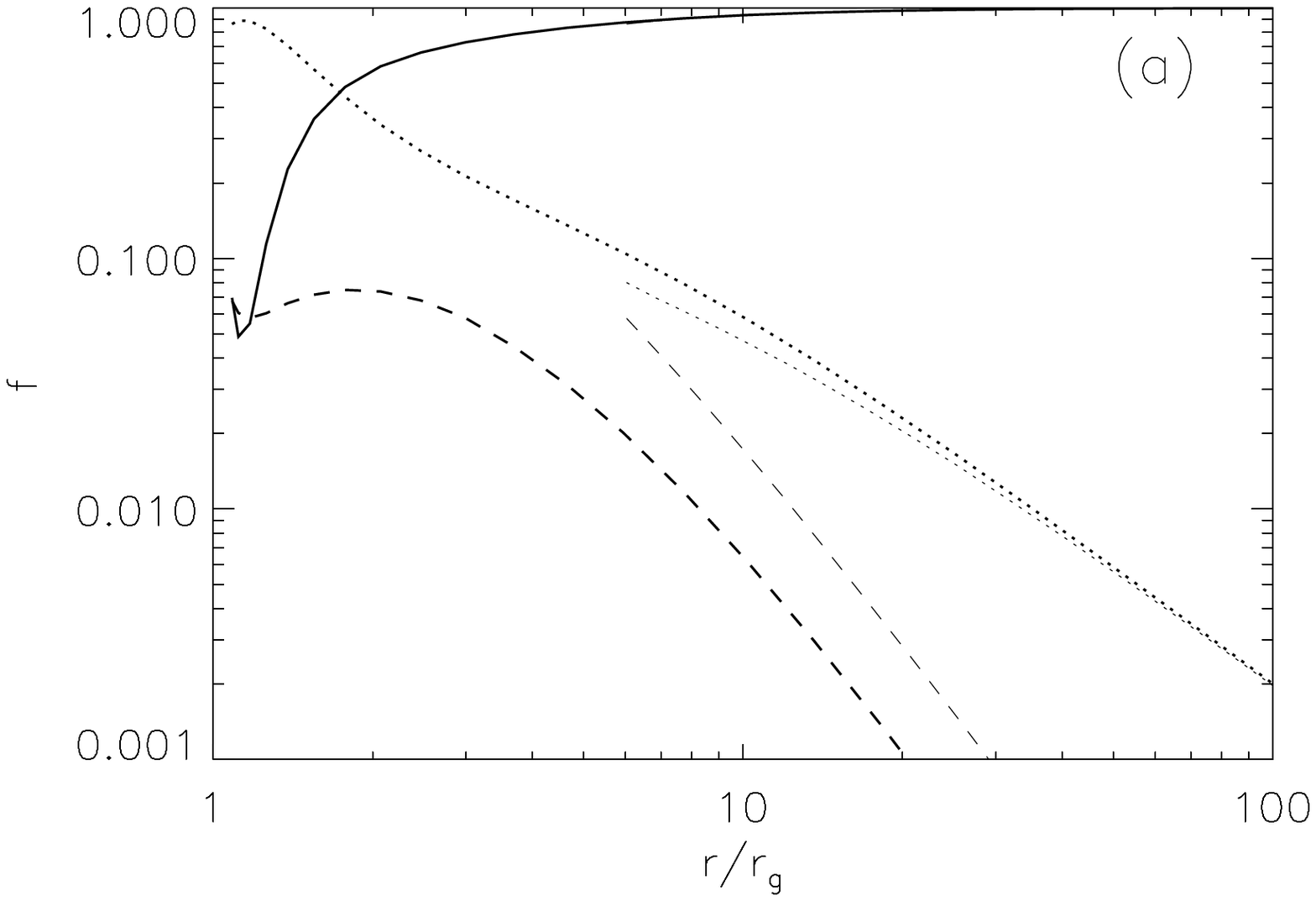,width=3.6in}} 
\centerline{\psfig{file=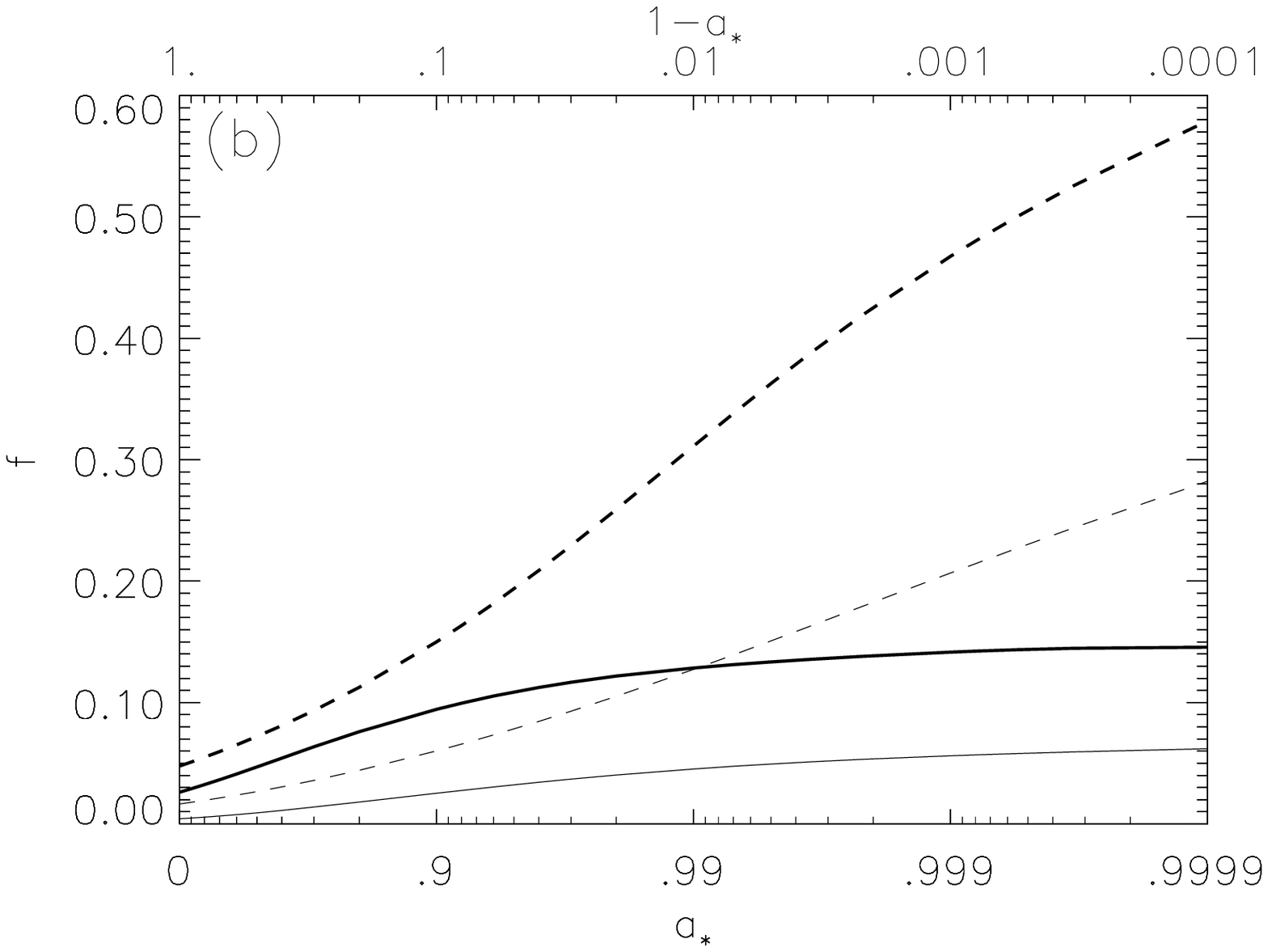,width=3.6in}} 
\noindent{
\scriptsize \addtolength{\baselineskip}{-3pt}
\vskip 1mm
\begin{normalsize}
Fig.~2.\ (a) Fraction of locally emitted flux which reaches infinity 
(solid lines), returns to the disk (dotted lines), and enters the black hole 
or returns inside $r_{ms}$ (dashed lines) versus radius.  The heavy lines
are for $a_*=0.9999$ and the lighter lines for $a_*=0$. (b)
Fraction of energy that returns to accretion disk integrated over radius
(dashed lines) or that is absorbed by the black hole (solid lines) as a
function of spin.  The heavy lines are for an infinite efficiency
disk, while the lighter lines are for a Novikov-Thorne disk.
\end{normalsize}
\vskip 3mm
\addtolength{\baselineskip}{3pt}
}

     The enhanced dissipation near $r_{ms}$ also leads to an increase
in the fraction of captured photons.  We have computed this fraction
integrated over all radii, $f_{BH}$, 
using the same general relativistic transfer code just described.
Our results for this effect are also illustrated in Figure 2b, again for 
$\epsilon = \epsilon_0, \infty$.  The $\epsilon = \epsilon_0$ results
agree with Thorne (1974).  Equation [\ref{fraction}] applies to 
$f_{BH}$ as well.  The fraction of locally generated radiation that ultimately 
escapes from the disk to infinity is simply $f_{esc} \equiv 1-f_{BH}$.
Radiation that returns to the accretion disk we assume is reradiated
isotropically and locally, and thus eventually reaches infinity or the
black hole.  We fold these multiply reprocessed photons into the final
result.  The nominal accretion efficiency, $\epsilon$, is then multiplied 
by $f_{esc}$ to find the actual radiative efficiency of the flow.  The
largest $f_{BH}$ is 0.15, achieved for $\epsilon \rightarrow \infty$ and
$a_* \rightarrow 1$.

The black hole bends the radiation back to the disk so that an observer 
on the disk sees the far side of the disk as a mirage above the black hole, 
which peaks in brightness within a few $r_g$ of the disk plane.  The flux 
at large radius then scales as $H/r^{-3}$, where $H$ represents the flux-weighted 
height of the image above the disk plane.  The ratio of the returning
radiation to locally generated radiation, $R_{ret}(\epsilon,a_*,r)$,
varies as a function of radius.  For a Novikov-Thorne disk, $R_{ret}$ 
is infinite at $r_{ms}$, then decreases rapidly, asymptoting
to a constant for $r \gtrsim 10 r_g$.  In the case of an infinite
efficiency disk (with $\dot M_o=0$), the locally generated surface
brightness scales as $r^{-7/2}$ at large radius, while the returning
radiation scales as $r^{-3}$, so $R_{ret}$ diverges as $r^{1/2}$ at large
radius.  For finite $\Delta \epsilon$, the returning flux may dominate
at intermediate radii; however, at large radius, $R_{ret}$ asymptotes to
a constant due to the fact that for large enough radius both returning 
and locally generated flux scale as $r^{-3}$.  For $r \gtrsim 10 r_{ms}$ 
and $\epsilon \le 1$, $R_{ret}$ differs by at most 25\% from the value 
at $r=\infty$.  We computed $R_{ret}(\epsilon,a_*,\infty)$ as a function 
of $a_*$ and $\epsilon$; this function is shown in Figure 3.
Fitting formulae for this quantity are given in the appendix;
these formulae can be used to compute the returning
flux at large radius for arbitrary $a_*, \epsilon$.
As can be seen in Figure 3, the returning radiation can
be a significant fraction of the locally generated radiation, and may
therefore be important for construction of
disk atmospheres.  Returning radiation can also lead to
significant fluctuations on the light-crossing time, as discussed in \S 3.3.
\vskip 2mm
\hbox{~}
\centerline{\psfig{file=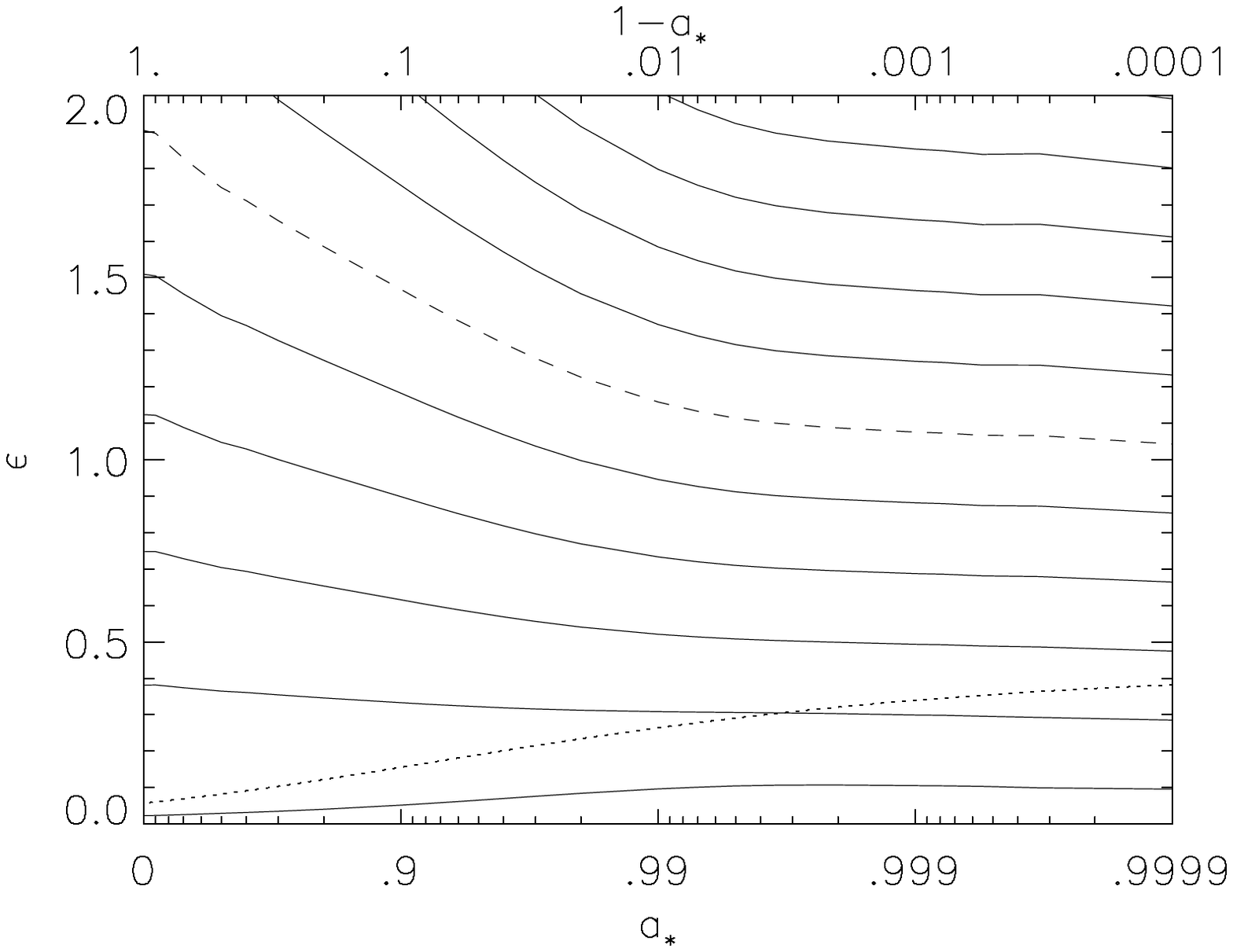,width=3.6in}} 
\noindent{
\scriptsize \addtolength{\baselineskip}{-3pt}
\vskip 1mm
\begin{normalsize}
Fig.~3.\  Contour plot of $R_{ret}$ at large radius.  Solid contours are
shown with spacing of 0.2, from $R_{ret}=0$ at the bottom to $R_{ret}=1.8$
at the top.  The dotted line is $\epsilon_0$; the dashed line  is the
contour of $R_{ret} = 1$.
\end{normalsize}
\vskip 3mm
\addtolength{\baselineskip}{3pt}
}

\section{Consequences}

\subsection{Black hole growth and spin-up (or spin-down)}

     Accreting matter enters the black hole with a certain amount of
angular momentum, changing the spin of the black hole.  When there is no
stress at the marginally stable orbit, the angular momentum absorbed
per unit rest mass accreted is exactly the specific angular momentum
of the marginally stable orbit, $L^\dagger_{ms} = M F_{ms} C_{ms}^{-1/2} x_{ms}^{1/2}$
(here we use conventional relativistic units in which $G = c = 1$).
However, when there are stresses at $r_{ms}$, angular momentum is transferred
from the matter inside $r_{ms}$ to the disk.  This reduces the accreted
angular momentum by an amount ${\cal L}_{ms} L^\dagger_{ms}$, where
\begin{equation}
{\cal L}_{ms} =  x_{ms} B_{ms}C_{ms}^{1/2} F_{ms}^{-1} \Delta\epsilon 
\end{equation}
when all the energy liberated in the plunging region is delivered to
the disk in the form of work done by torque.
${\cal L}_{ms} = 3\sqrt{2} \Delta\epsilon $ when $a_* = 0$, falling 
towards $\sqrt{3} \Delta\epsilon$ when $a_*$ approaches one.  Thus, the
rate at which black holes are spun up is substantially reduced relative
to what would be expected in the conventional picture.  
Surprisingly, even when the black hole is initially spinless, it can 
be spun {\it backwards} when $\epsilon > 1-1/\sqrt{2} \sim 0.29$!

    Considerations of black hole spin-up also place an upper bound on
the possible increase in efficiency due to torques on the disk.  By the
second law of black hole dynamics, the area, $A$, of the black hole 
must increase with time; that is,
\begin{equation}
{dA \over dt} = {\partial A \over \partial M} {d M \over dt } +
{\partial A \over \partial J} {dJ\over dt} > 0.
\end{equation}
Since $dM/dt$ and $dJ/dt$ both depend on $a_*$ and $\epsilon$, this
constraint can be changed into a constraint on $\epsilon$ as a function
of $a_*$.  For $a_* < 0.3584$, there is a maximum achievable efficiency
\begin{equation}
\epsilon_{max} = 1-{a_* C_{ms}^{1/2} x_{ms}^{3/2} \over a_*^2 + a_* 
x_{ms}^{3/2} - 2 (1+\sqrt{1-a_*^2})}.
\end{equation}
Note that $\epsilon_{max}=1$ for $a_*=0$, because, of course,
there is no spin energy to tap.  For $a_* \simeq 0.3584$, the denominator
equals zero, so $\epsilon_{max}$ diverges;  above this critical 
spin, the decrease in angular momentum dominates the change in surface
area, eliminating any upper bound on $\epsilon_{max}$.  
When accreted radiation is included, $\epsilon_{max}$ increases slightly.
We plot $\epsilon_{max}$ in Figure 4.

\vskip 2mm
\hbox{~}
\centerline{\psfig{file=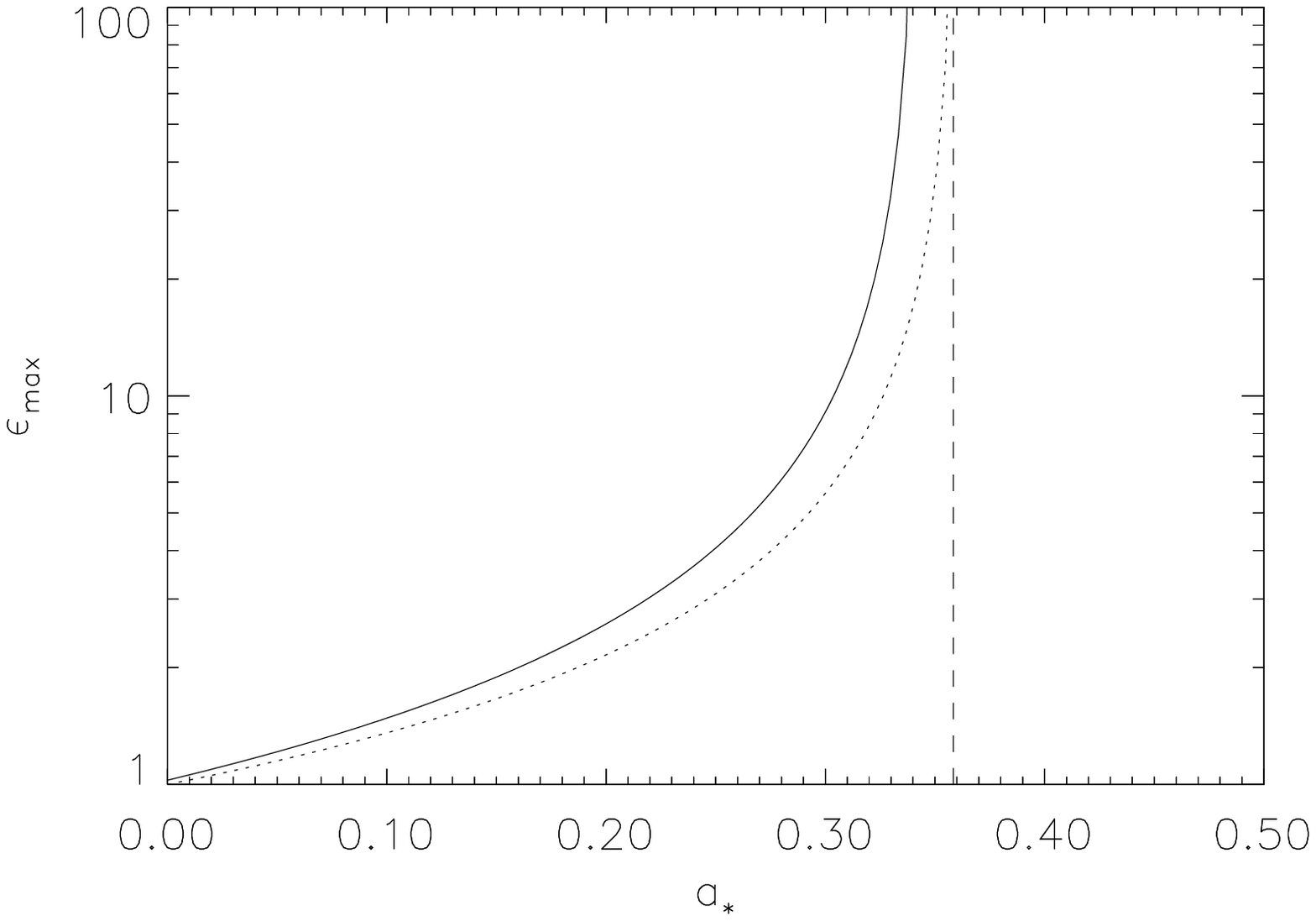,width=3.6in}} 
\noindent{
\scriptsize \addtolength{\baselineskip}{-3pt}
\vskip 1mm
\begin{normalsize}
Fig.~4.\  Plot of the maximum achievable efficiency ($\epsilon_{max}$) vs.
spin due to the limit imposed by the second law of black hole dynamics
with (solid) and without (dotted) the effects of radiation.
\end{normalsize}
\vskip 3mm
\addtolength{\baselineskip}{3pt}
}
    As Thorne (1974) showed, when $a_*$ approaches one, the angular momentum
of the black hole is also affected by photon capture.  Most of the photons
emitted close to $r_{ms}$ directed against the sense of black hole rotation
are captured by the black hole, whereas fewer of the prograde photons fall
into the hole.  This extra negative angular
momentum prevents it from spinning up all the way to $a_* = 1$.
Under the assumption of isotropic radiation in the fluid frame (and, of
course, zero stress at $r_{ms}$), Thorne
estimated that the maximum achievable $a_* \simeq 0.998$.  To describe
this effect in our context, we again normalize to $L^\dagger_{ms}$, so
that the photon ``reverse torque" per unit accreted
mass is ${\cal L}_\gamma \equiv -(L^\dagger_{ms}\dot M_o)^{-1}
(dJ/dt)_{rad}$, where the notation is adapated
from Thorne (1974).
 
    Combining the effects of mechanical torque and photon capture, we
find that the net rate of change of the black hole's angular momentum
is 
\begin{equation}
{dJ \over dt} = L^\dagger_{ms} \dot M_o (1 - {\cal L}_{ms} - {\cal L}_\gamma).
\end{equation}
Because both ${\cal L}_\gamma$ and ${\cal L}_{ms}$ depend on the state of
magnetic coupling, as well as on $a_*$, it is no longer possible to speak
of a definite upper bound on the attainable black hole spin.  Rather,
one can instead define the accretion efficiency, $\epsilon_{eq}(a_*)$, at
which $d a_* / dM = 0$ for a given $a_*$;  for $\epsilon >
\epsilon_{eq}$, the black hole is spun down due to accretion.  
To compute $\epsilon_{eq}$, we write $-L_\gamma = (dJ/dt)_{rad}/
(\dot M_o L^\dagger_{ms}) = J_1' + \epsilon J_1'$ and 
$(dM/dt)_{rad}/\dot M_o  = M_1' + \epsilon M_2'$  (we give fitting
formulae for these functions in the appendix).
Then, the equilibrium efficiency is given by:
\begin{equation}
\epsilon_{eq} = {2a_*(1+M_1')-\epsilon_0x_{ms}^{3/2}B_{ms}-L^\dagger_{ms}(1+J_1')
\over 2a_*(1-M_2') -x_{ms}^{3/2}B_{ms} + L^\dagger_{ms} J_2'}.
\end{equation}
\vskip 2mm
\hbox{~}
\centerline{\psfig{file=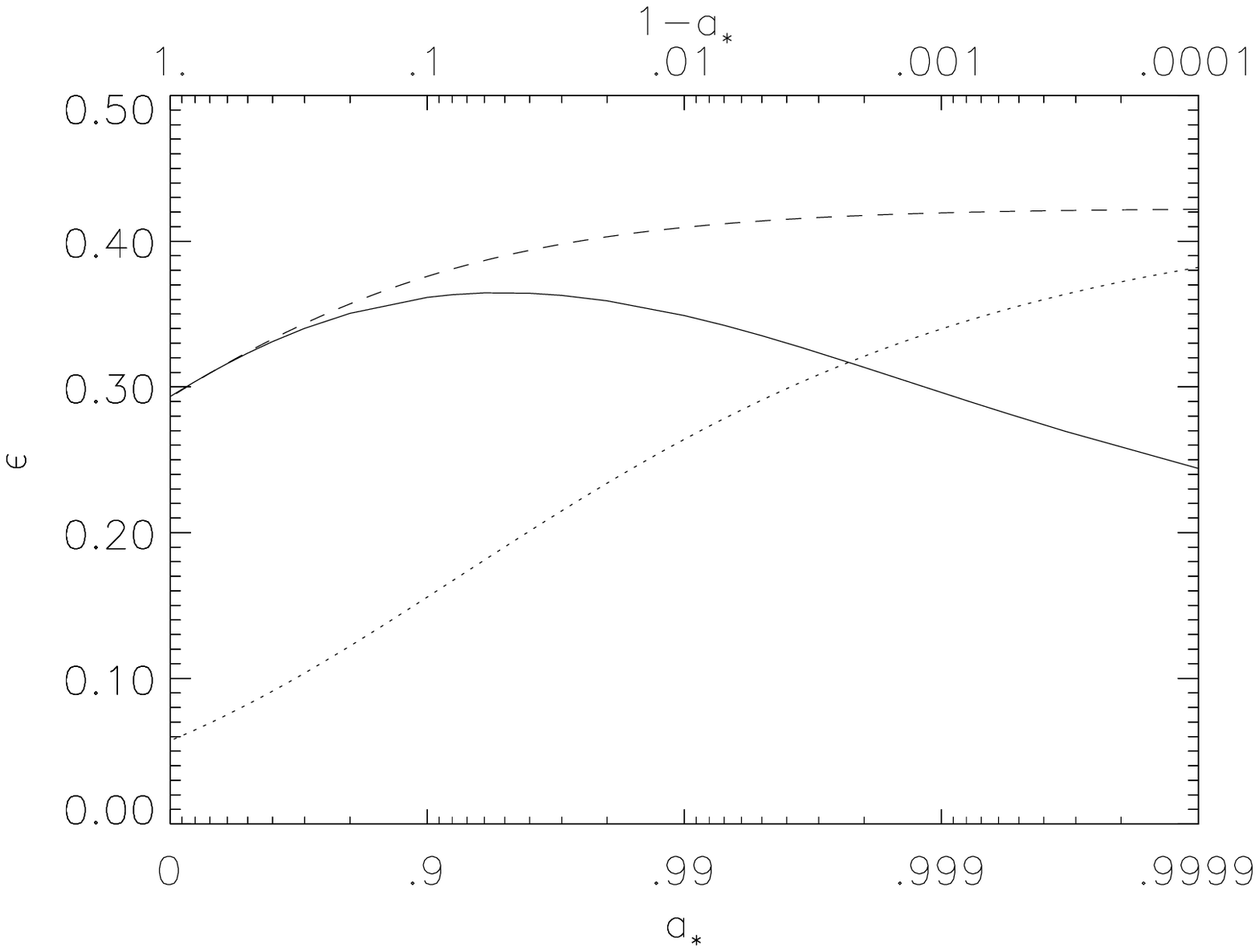,width=3.6in}} 
\noindent{
\scriptsize \addtolength{\baselineskip}{-3pt}
\vskip 1mm
\begin{normalsize}
Fig.~5.\  Plot of $\epsilon_{eq}$ with (solid) and without (dashed)
effects of returning radiation.  The dotted line is $\epsilon_0$,
the Novikov \& Thorne efficiency.
\end{normalsize}
\vskip 3mm
\addtolength{\baselineskip}{3pt}
}
Figure 5 shows
$\epsilon_{eq}(a_*)$ when only magnetic and matter torques are included (dashed
line), and when magnetic, matter, and radiation torques are included (solid
line), as well as $\epsilon_0(a_*)$, the efficiency of accretion (not in
equilibrium) when magnetic torques are ignored (dotted line).
When accreted radiation is ignored, equation [17] simplifies to:
\begin{equation}
\epsilon_{eq} =  1 - { \sqrt{C_{ms}} \over 2 - B_{ms}}.
\end{equation}
This limit is accurate for $a_* < 0.5$, and only creates a significant
error for $a_* > 0.9$, for the radiation
torque is unimportant when the spin is relatively small.
Some interesting limiting values are $\epsilon_{eq}(0)= 1-1/\sqrt{2}$, and,
when radiation effects are ignored, $\epsilon_{eq}(1)= 1-1/\sqrt{3}$, which
equals $\epsilon_0(1)$.  However, when the radiation torque is included,
$\epsilon_{eq}$ is significantly reduced for $a_* >  0.9$, and
$\epsilon_{eq} < \epsilon_0$ for $a_* > 0.998$, the same maximum spin
found by Thorne (1974).
The maximum equilibrium efficiency is 0.36, and occurs at $a_* = 0.94$.

    A variety of spin histories is possible in this picture, for the
efficiency is controlled jointly by the strength of the magnetic torques
and the black hole spin.  If the torque on the disk is always positive,
the region below the curve $\epsilon_0 (a_*)$ in figure 5
is unreachable.  In
that case, $a_* = 0.998$ would still be the maximum spin achievable by
accretion, although other spin-up mechanisms, such as black hole mergers
or non-magnetic accretion, might permit this limit to be exceeded.

\vskip 2mm
\hbox{~}
\centerline{\psfig{file=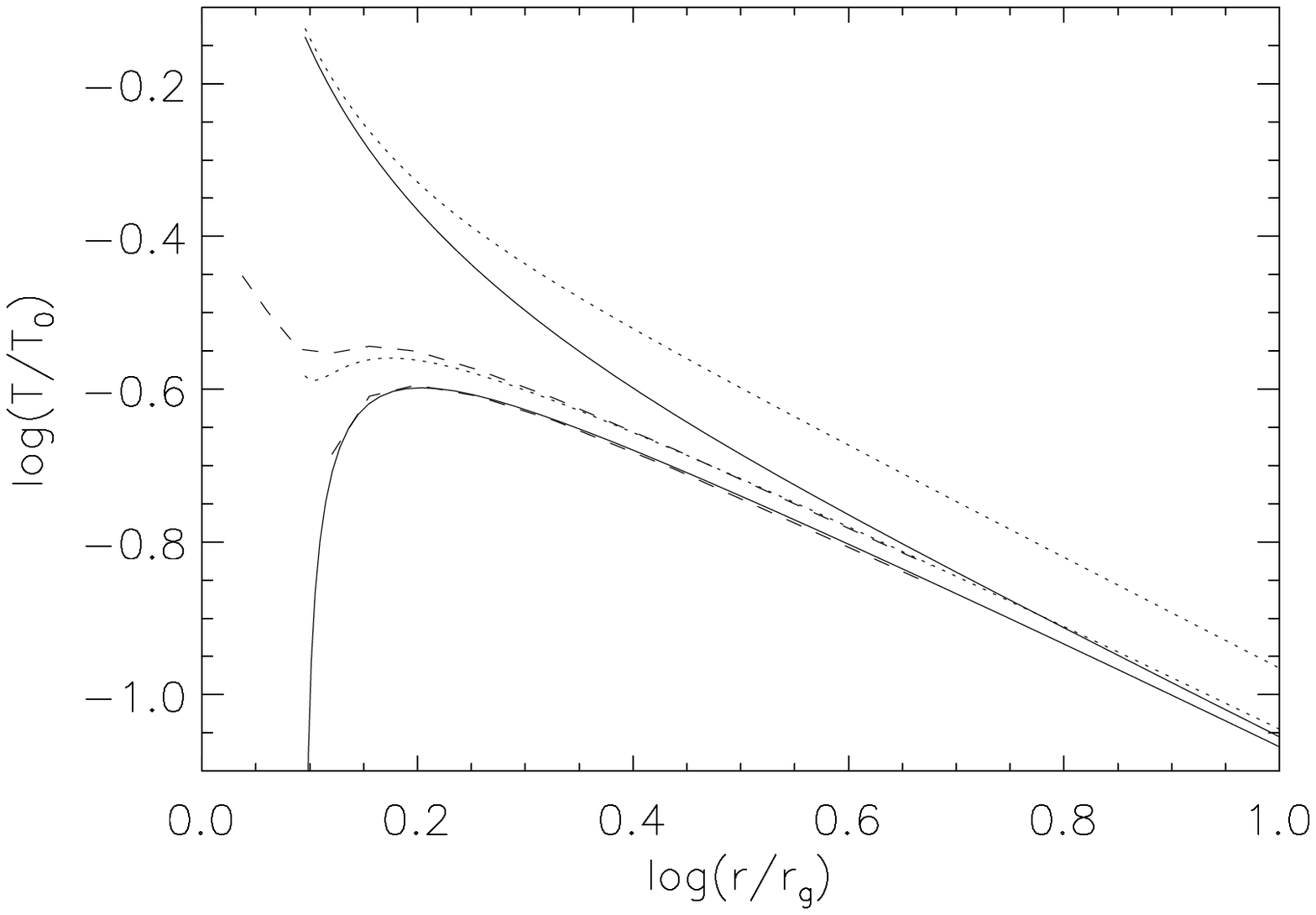,width=3.6in}} 
\noindent{
\scriptsize \addtolength{\baselineskip}{-3pt}
\vskip 1mm
\begin{normalsize}
Fig.~6.\  Plot of the effective temperature vs. radius for a black hole
with $a_*=0.998$.
The temperature is normalized to $T_0 \equiv [\dot M_o c^2/(r_g^2\sigma_B
)
]^{1/4}$, where $\sigma_B$ is the Stefan-Boltzmann constant.
The lower three curves are for a Novikov-Thorne disk, while the upper
two are for an $\epsilon = 1$ disk.  The solid lines are without
returning radiation, while the dotted lines include returning radiation.
The dashed line shows the result of Cunningham (1976).
\end{normalsize}
\vskip 3mm
\addtolength{\baselineskip}{3pt}
}
\subsection{Emitted spectrum}

The effective temperature is determined by the sum of the locally
generated and returning flux.  Figure 6 illustrates the effects discussed 
in \S 2, showing both how the intrinsic dissipation varies as a function of 
radius when there is a torque on the disk inner edge, and the total surface 
flux if one assumes that any incident radiation is absorbed.  By comparing
the curves for $\Delta\epsilon = 0$ with the other curves, it is clear
that the additional stress has two effects: the additional intrinsic dissipation
creates a region at small radius where the effective temperature is rather
higher than the disk could achieve otherwise; and returning radiation
elevates the effective temperature at all radii, especially when $a_*$
is near unity and $\Delta\epsilon \sim 1$ or more.

     Ideally, detailed atmosphere calculations should be performed in order to
ascertain the predicted disk spectrum, 
\vskip 2mm
\hbox{~}
\centerline{\psfig{file=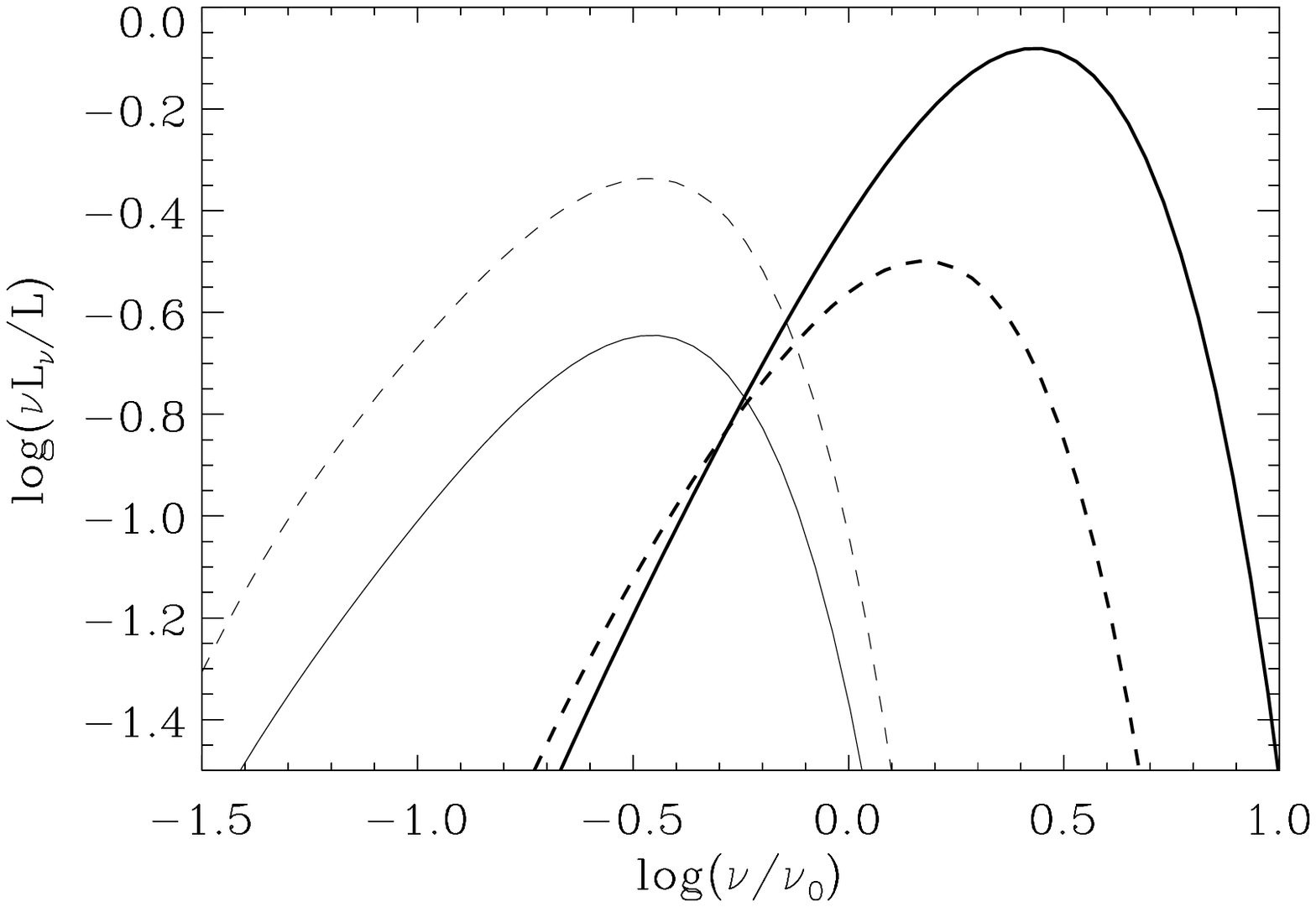,width=3.6in}} 
\noindent{
\scriptsize \addtolength{\baselineskip}{-3pt}
\vskip 1mm
\begin{normalsize}
Fig.~7.\  Comparison of the spectra as a function of inclination
angle for $\epsilon=\epsilon_0$ (dashed lines) and $\epsilon = \infty$
(solid lines).  The other parameters are $a_*=0.998$,
and $r_{out}=500 r_g$.  The heavy lines are for $\mu=0.01$ while
the lighter lines are for $\mu=0.99$.  The frequency is scaled
to $\nu_0 \equiv (k/h)\left[L/(r_g^2 \sigma_B)\right]^{1/4}$.
The quantity $L_\nu \equiv 4\pi D^2F_\nu(\mu)$, where $F_\nu(\mu)$
is the flux seen by a Euclidean observer at distance $D$ and angle
$\mu=\cos{i}$ relative to the accretion disk.
\end{normalsize}
\vskip 3mm
\addtolength{\baselineskip}{3pt}
}
\noindent with the downgoing flux of returning
radiation included in the upper boundary condition.  Interesting effects might 
well be expected due to comparable amounts of heat arriving from above as from 
below.  Pending the completion of that work, we make the much simpler assumption
that the intensity at the surface of the disk is a blackbody at the local
effective temperature, and isotropic in the outward half-sphere.  With
that assumption, Figures 7  and  8 show the predicted integrated spectrum for
a variety of values of $a_*$, $\Delta\epsilon$, and inclination (parameterized
by $\mu = \cos{i}$).   
\vskip 2mm
\hbox{~}
\centerline{\psfig{file=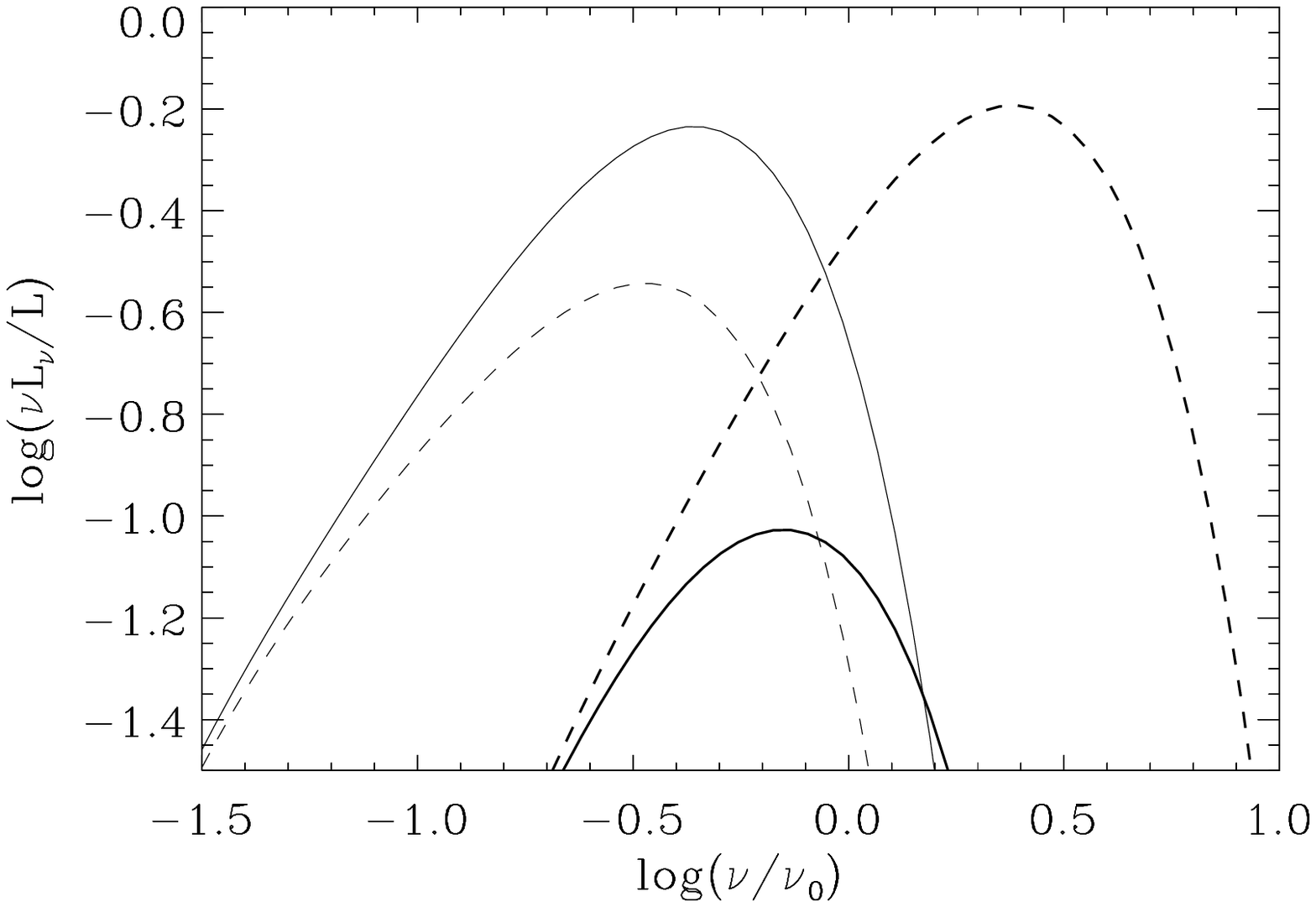,width=3.6in}} 
\noindent{
\scriptsize \addtolength{\baselineskip}{-3pt}
\vskip 1mm
\begin{normalsize}
Fig.~8.\  Comparison of the spectra as a function of inclination angle
for $a_*=0$ (solid lines) and $a_*=0.998$ (dashed lines) with
$\epsilon=1$ and $r_{out} = 500 r_g$.  The heavier lines are for
$\mu =0.01$, while the lighter lines for $\mu=0.99$.  Units are the
same as in Figure 7.
\end{normalsize}
\vskip 3mm
\addtolength{\baselineskip}{3pt}
}
\noindent Figure 7 shows that for fixed luminosity and
large spin, the efficiency of accretion can change the observed flux by 
factors of a few at different inclination angles.  The angle dependence
of the flux depends strongly on frequency---the highest frequency
radiation is concentrated towards the disk plane, while the lowest frequencies 
are radiated as $\cos{i}$.  
Figure 8 shows the dependence of the spectrum
on black hole spin for fixed luminosity and efficiency.  The relativistic effects
are much stronger for the higher spin, hardening the edge-on spectrum
and causing strong limb-brightening at the highest frequencies.  In contrast,
the disk around the Schwarzschild hole is limb-darkened at most frequencies,
and, when face-on, is brighter by a factor of a few at the mid-range 
frequencies than the extreme Kerr hole.
These effects may also impact the profiles of Fe K$\alpha$ emission
lines.  If their emissivity is proportional to the local flux,
the enhanced flux in the inner rings of the disk strengthens
the red wings of the lines when viewed more or less face-on.  We plot the
profiles of K$\alpha$ lines for disks with $\Delta\epsilon=0$ and
$\epsilon = \epsilon_{eq} = 0.293$  for $a_*=0$ and $i=30^\circ$ in Figure 9.
Disks with higher spin have a smaller change in the shape
of the iron line as a function of $\epsilon$ because the returning radiation is
much stronger and creates an emissivity profile very similar to the 
Novikov-Thorne profile.
Magnetized accretion may also lead to enhanced coronal activity immediately
above the plunging region (Krolik 1999); if so, this would provide a
physical realization for models like those of Reynolds \& Begelman (1997),
which call for a source of hard X-rays on the system axis a few gravitational
radii above the disk plane.
\hbox{~}
\centerline{\psfig{file=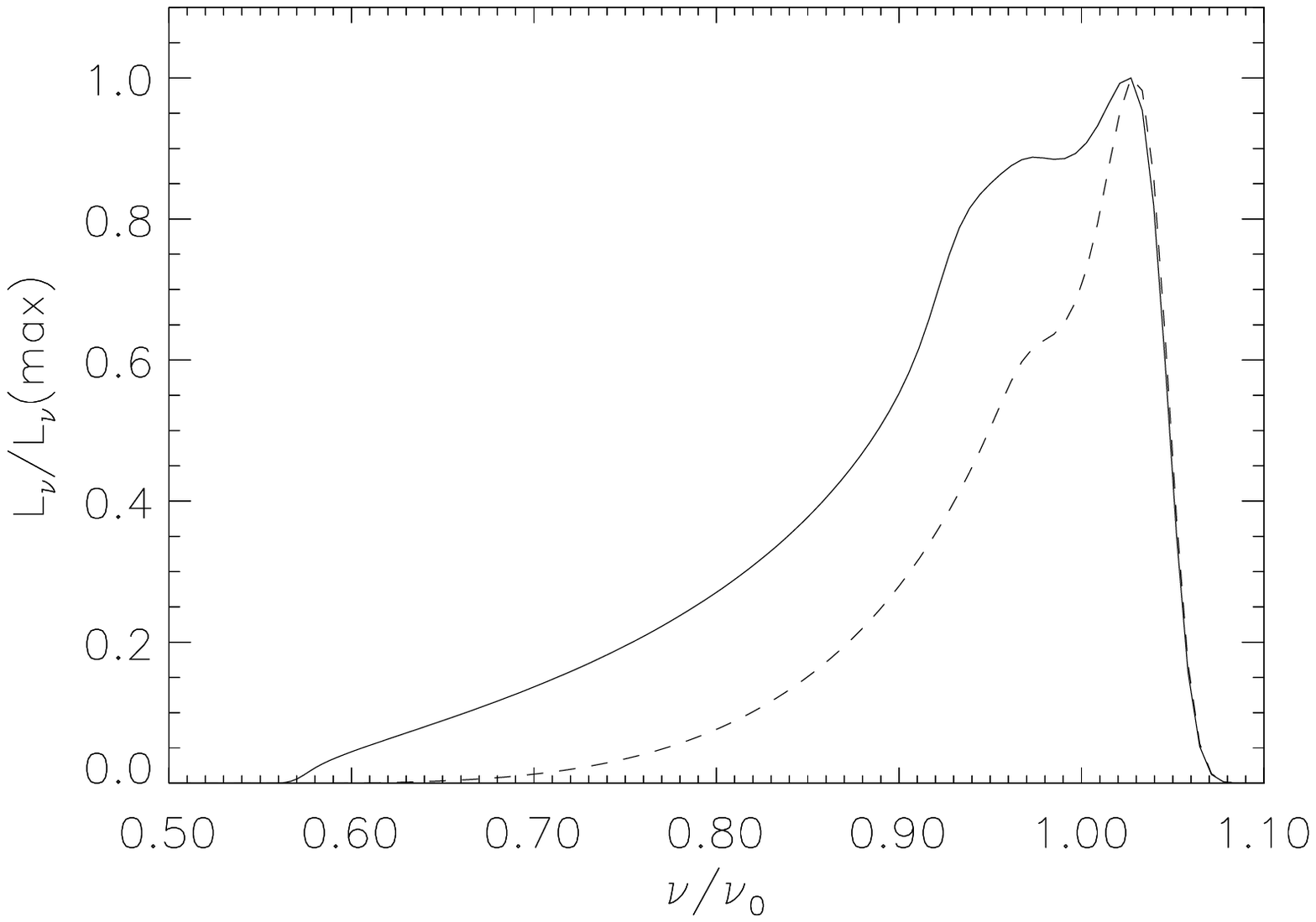,width=3.6in}} 
\noindent{
\scriptsize \addtolength{\baselineskip}{-3pt}
\vskip 1mm
\begin{normalsize}
Fig.~9.\ Profiles of Fe K$\alpha$ lines for $a_*=0$, $i=30^\circ$, and
$\epsilon = \epsilon_{eq} = 0.293$ (solid curve), $\epsilon = \epsilon_0$
(dashed curve).  Frequency is normalized to unshifted line frequency,
and line amplitude is normalized to the line maximum. \\
\end{normalsize}
\addtolength{\baselineskip}{3pt}
}

\subsection{Coordinated variations}

      When $\Delta\epsilon$ is comparable to the ordinary efficiency, the
inner rings of the disk radiate an amount of energy comparable to that
radiated by all the rest of the disk.  When, in addition, $a_*$ is large
enough that $f_{\rm ret}$ is significant, much of the light produced
even at larger radii is reprocessed energy from the additional dissipation.
If there are variations in that dissipation rate, they will be reproduced---at
appropriate delays---in the reprocessed light.  A prediction of this
picture is therefore that fluctuations at a wide range of frequencies $\nu$
should all be describable as driven by a single source.  When the
fluctuations in the returning flux are small compared to the mean local
flux (combining both the intrinsic and the mean returning flux), the
relation between input and output may be written as the linear convolution
\begin{equation}
\delta L_\nu (t) = \int \, d\tau \Psi_\nu (\tau) \delta L_c (t - \tau),
\end{equation}
where $\delta L_c (t)$ is the history of fluctuations in the
intrinsic output near $r_{ms}$
and $\Psi_\nu$ is a frequency-specific ``response function" \footnote{ We use
the term ``response function" to avoid confusion with the relativistic
``transfer function".} that describes
the distribution of relevant light-travel times.  Note, however, that
if there is a corona at small radii that receives a significant fraction
of the total dissipation (indeed, such a corona might receive much
of the extra accretion energy: Krolik 1999), it will also drive fluctuations
in the output of the outer disk in very much the same manner, and with
a substantially identical response function.

   The response function $\Psi_\nu(\tau)$ is also predicted by this model.
To compute this function we make several simplifying approximations: that
all the returning radiation is absorbed; that it is reradiated in a spectrum
that is locally blackbody and isotropic in the outer half-sphere; and that 
the radii of interest are far enough out in the disk that relativistic 
effects may be ignored.  Then
\begin{eqnarray}
\Psi_\nu \simeq {f_{\rm ret}\mu r_* h\nu^3\over 2 c L_*} \int_{r_1}
^{r_2}
\, dr \,
{r^{3/4} \over e^{r^{3/4}} + e^{-r^{3/4}} - 2}\cr
\times
\left[ 1 - \mu^2 - \left( c\tau/r - 1\right)^2 \right]^{-1/2} ,
\end{eqnarray}
where radius $r$ and $c\tau$ are measured in units of $r_*$, the radius at
which $h\nu = kT$ when the flux takes its mean value,
$r_{in}$ is the innermost radius at which the returning flux
is $\propto r^{-3}$, $r_1 = max[r_{in},c \tau /(1-\sqrt{1-\mu^2})]$, $r_2
= c\tau /(1-\sqrt{1-\mu^2})$, and $\mu$ is the cosine of
the inclination angle.  The characteristic radius $r_*$ is given by
\begin{equation}
r_* = \left({L_* r_{in} k^4 \over 4\pi \sigma h^4 \nu^4}\right)^{1/3},
\end{equation}
where $k$ is the Boltzmann constant, $\sigma$ is the Stefan-Boltzmann
constant, and $L_*$ is the mean value of the luminosity emitted by the
portion of the disk whose emissivity is $\propto r^{-3}$. 

    Some sample response functions are illustrated in Figure 10.  All of the
curves have significant tails extending out to $\sim 10 r_*/c$, and,
almost independent of inclination angle, the ``half-response time" (in the
sense of $\int d\tau \Psi_\nu$) occurs at $\tau \simeq 3 r_*/c$.
However, the peak in the response function becomes sharper
and moves to smaller multiples of $r_*/c$ as the inclination angle
increases.  Two effects account for this behavior.
The tails are due to the fact that the temperature declines only
as $r^{-3/4}$, so that the Wien cut-off sets in relatively slowly.  The
sharp peaks at small lag exhibited by disks with larger inclination angle
are due to the significant amount of disk surface that lies close to
the line of sight for those viewing angles.
\vskip 2mm
\hbox{~}
\centerline{\psfig{file=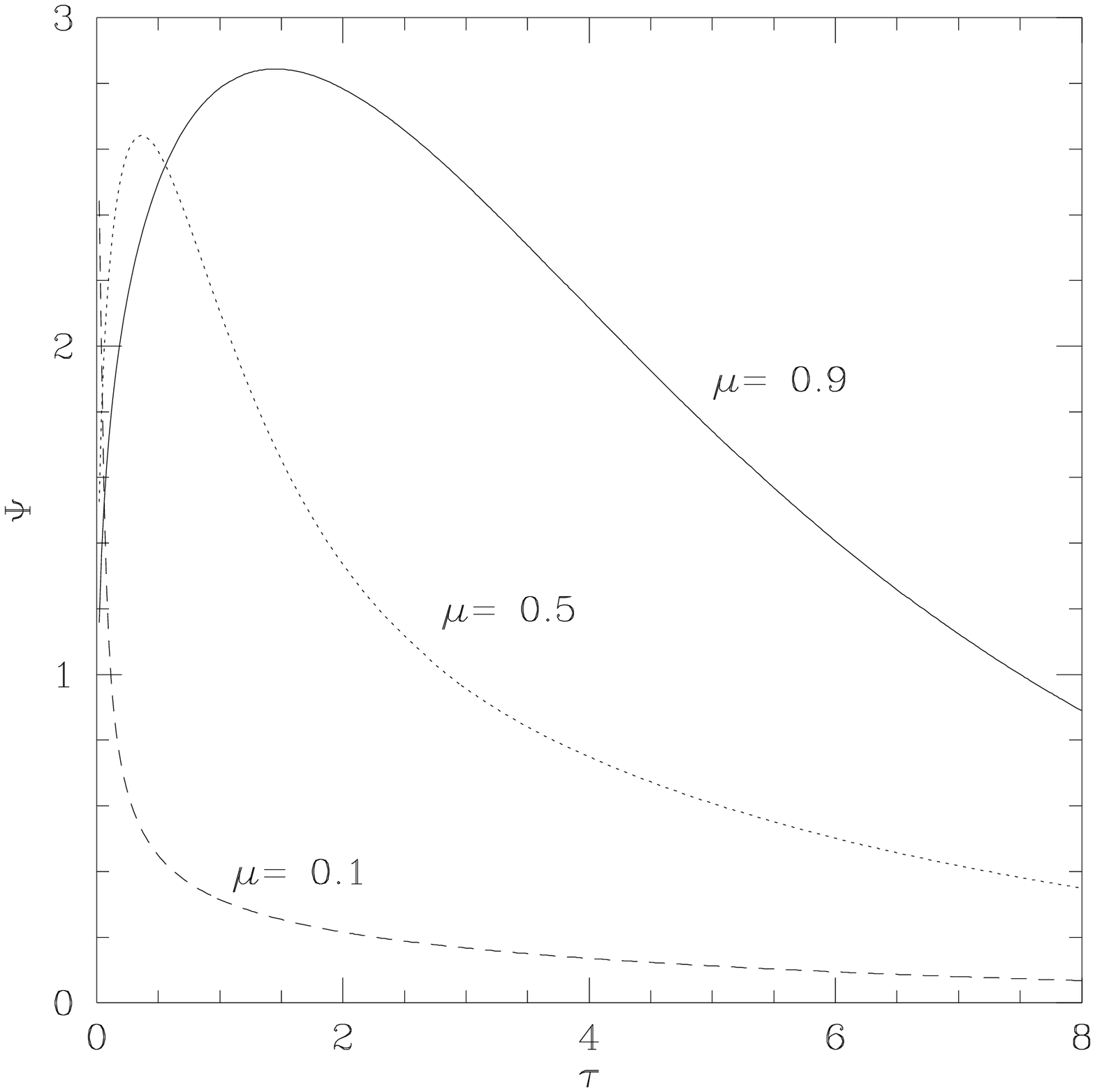,width=3.6in}} 
\noindent{
\scriptsize \addtolength{\baselineskip}{-3pt}
\vskip 1mm
\begin{normalsize}
Fig.~10.\  Plot of continuum response $\Psi_\nu$ as a function of lag
for different inclinations. $\Psi_\nu$ is in units of $f_{\rm ret}r_*
h\nu^3/(2 c L_*)$ and $\tau$ is in units of $r_*/c$.
\end{normalsize}
\vskip 3mm
\addtolength{\baselineskip}{3pt}
}

\subsection{Polarization}

    In fact, the inner rings of realistic disks, whether in AGN or
Galactic black holes, are likely to be scattering-dominated, so that their
albedo to the returning radiation will be significantly greater than zero.
The scattered light may then be polarized.

    As discussed in \S 2.2, the maximum altitude $H$ above the disk plane 
achieved by any photon
that ultimately returns to the disk cannot be much greater than a few
gravitational radii.  If the disk flare is small (see \S 3.6 for further
discussion), the returning photons striking the disk at
radius $r$ must then arrive from an angle $\simeq H/r \ll 1$ from the
disk equator.  When electron scattering is the dominant scattering opacity
(as is nearly always the case), only those photons polarized parallel to the
disk normal can scatter to outgoing directions near the equatorial plane
but perpendicular to the original photon direction.
The result is that disks viewed obliquely should acquire
a small amount of polarization parallel to the disk axis, especially at 
the high frequencies produced predominantly in the inner rings.

    To quantify this suggestion, we have computed the disk spectrum,
treating the locally generated radiation as a blackbody, and assuming 
the returning radiation is scattered off a semi-infinite electron 
scattering atmosphere (Chandrasekhar 1960).  We assume the locally
generated disk flux has either (1) the polarization of a semi-infinite 
electron scattering atmosphere (Chandrasekhar 1960) or (2) is unpolarized.  
The true polarization will
be modified by Faraday rotation due to magnetic fields in the disk's
atmosphere and absorption/emission (Agol, Blaes, \& Ionescu-Zanetti 1998), 
but the true answer will likely lie between our two assumptions.  Figure 11 
shows the flux, polarization, and polarization angle computed under these two 
assumptions.  The spectrum is much broader than would be predicted by
complete absorption, and returning radiation can
cause a sharp rise in polarization towards the highest frequencies.  There is
a rotation in the polarization angle since the scattered returning radiation
tends to be polarized perpendicular to the disk plane,
while the locally generated
radiation is polarized parallel to the disk plane.  Whichever component
dominates the flux at a given frequency determines the strength and
angle of the polarization.  We have included all relativistic effects that
modify the final polarization angle (Laor, Netzer, \& Piran 1990, Agol 1997).
The returning fraction is largest for photons generated in the inner region
of the disk; consequently, the highest frequencies have the
largest scattered fraction and thus the highest polarization.  In addition,
the inner parts of the disk are strongly blueshifted, and the returning
radiation is (weakly) Compton up-scattered by the bulk motion of the disk.
\vskip 2mm
\hbox{~}
\centerline{\psfig{file=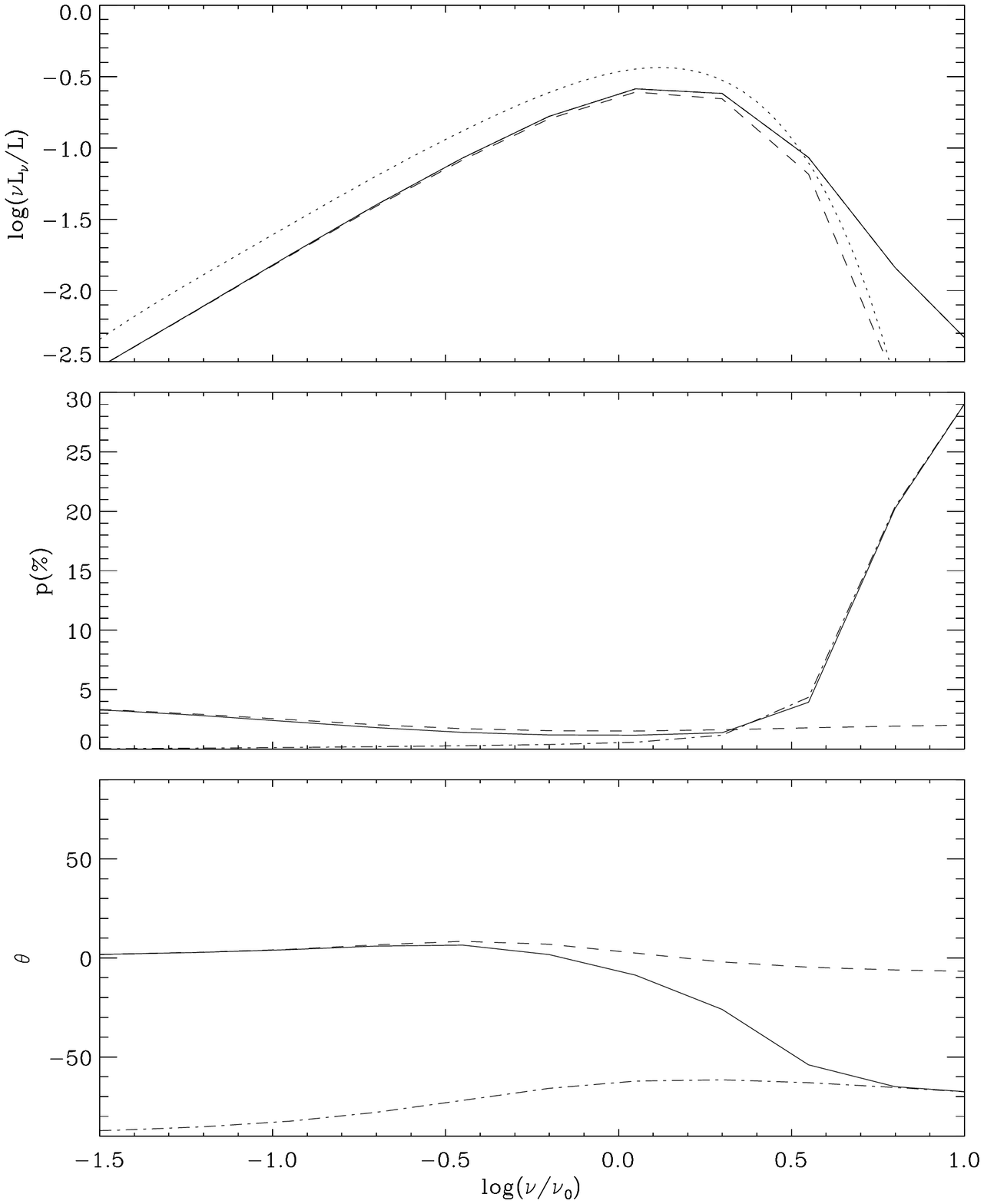,width=3.6in}} 
\noindent{
\scriptsize \addtolength{\baselineskip}{-3pt}
\vskip 1mm
\begin{normalsize}
Fig.~11.\  Flux, polarization, and polarization angle as a function
of frequency for a disk viewed with an inclination of $\mu=\cos{i}=0.2$,
with $a=0.998$, $\Delta\epsilon = 1$.  The dashed curves are for no
returning radiation;
dashed-dot for returning radiation, but unpolarized locally generated
flux; the solid curves are for returning radiation plus polarized
locally generated flux; and the dotted line in the top panel shows the
flux computed assuming complete absorption, as in \S 3.2.  The polarization
angle, $\theta$, is zero for ${\bf E}$ parallel to the disk plane.
The units are defined in Figure 7.
\end{normalsize}
\vskip 3mm
\addtolength{\baselineskip}{3pt}
}

\subsection{Bolometric Limb-brightening}

For a Newtonian disk, foreshortening causes limb-darkening proportional
to $\mu = \cos{i}$, where $i=0$ is a face-on disk. 
Relativistic effects cause beaming and bending of the radiation 
towards the equatorial plane, which decrease the limb-darkening for
disks around black holes.  For large $a_*$ and $\epsilon$, the relativistic
effects become so strong that a disk can actually become {\it limb-brightened}. 
In Figure 12 we show the bolometric disk flux as a function
of inclination angle for the cases $\epsilon = \infty$ (dashed line),
$\epsilon = 1$ (dotted line), and $\epsilon = \epsilon_0$ (solid line)
for $a_*=0.998$.  The limb-brightening is also be dependent on frequency
as shown in \S 3.2; in practice, determining this quantitatively will require 
a detailed disk atmosphere model.
\vskip 2mm
\hbox{~}
\centerline{\psfig{file=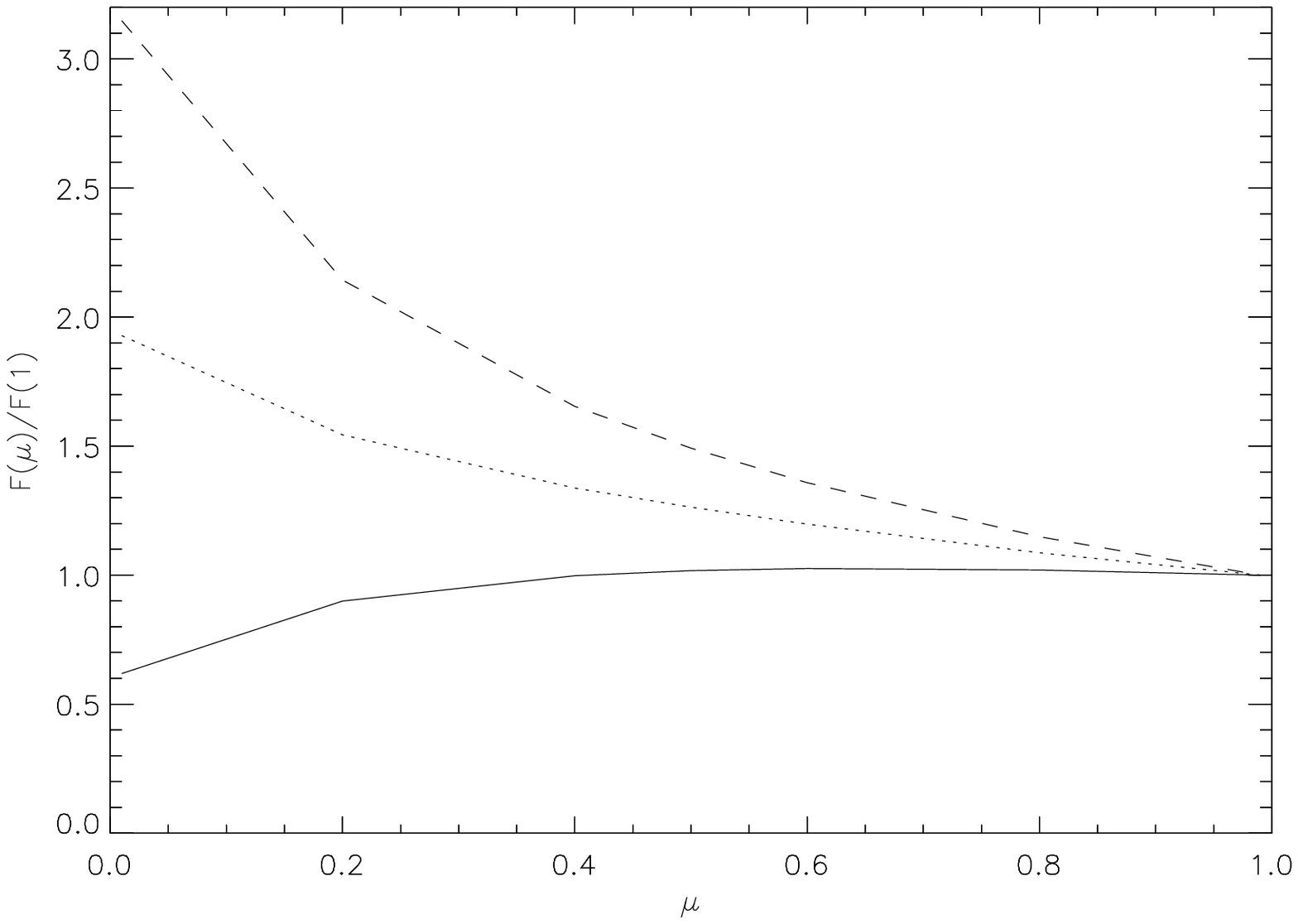,width=3.6in}} 
\noindent{
\scriptsize \addtolength{\baselineskip}{-3pt}
\vskip 1mm
\begin{normalsize}
Fig.~12.\  Comparison of the bolometric limb-brightening
of a disk for $a_*=0.998$
with $\epsilon = \epsilon_0$ (solid line), $\epsilon = 1$ (dotted line),
and $\epsilon = \infty$ (dashed line).  The inclination angle, $\mu
= \cos{i}$ is edge-on for $\mu=0$ and face-on for $\mu=1$.
\end{normalsize}
\vskip 3mm
\addtolength{\baselineskip}{3pt}
}

\subsection{Geometrical thickness of the disk}

     When the accretion rate is greater than a small fraction of the Eddington
rate, the innermost regions of accretion disks are expected to be supported
against the vertical component of gravity by radiation (Shakura \& Sunyaev
1973).  In that case, the disk's vertical thickness is directly proportional
to the ratio of the local radiation flux to the vertical component of
gravity; that is, $h \propto Fr^3/R_z$, where $R_z$ is the relativistic
adjustment to the vertical gravity (Page \& Thorne 1974; Abramowicz, Lanza,
\& Percival 1997).  At radii large enough that the relativistic effects are 
small, but not so large as to no longer be in the radiation-dominated regime 
in a Novikov-Thorne disk, $h$ should be constant.  In the relativistic
portion of the disk, $h$ would shrink $\propto R^{NT}/R_z$ if the stress at its
inner edge were zero; additional dissipation, depending on its strength, could actually make the disk
become somewhat thicker there (cf. equation 8).  We plot
some examples of $h(r)$ in Figure 13.  At the inner edge of the disk,
the height of the disk is non-zero when there is a non-zero torque, so the thin-disk approximation is valid only if $h(r_{ms}) \lesssim 0.1 r_{ms}$.
This criterion can be translated into
a limit on the extra luminosity due to the torque at the inner edge:
\begin{equation}
{L \over L_{Edd} } < 0.1 {2 \over 3 x_{ms}} \left(x_{ms}C_{ms}^{-1/2} F_{ms}^2 
-a^2 G_{ms} + a^2 C_{ms}^{1/2}\right) ,
\end{equation}
where $L = \Delta \epsilon \dot M_oc^2$.  This limiting luminosity
is plotted in Figure 14.
When $\Delta \epsilon = 0$, the thin disk approximation breaks down
if $h/r \gtrsim 0.1$ where $h/r$ is maximum.  This limit can in turn
be expressed as a limit on the luminosity $\epsilon_0 \dot M_o c^2$,
which is also shown in Figure 14.  The luminosity upper limit for the
infinite efficiency disk is much smaller than for the Novikov-Thorne
disk since $h/r$ peaks at $r_{ms}$, while for the 
Novikov-Thorne disk $h/r$ peaks at larger radius, where its magnitude is smaller
[$(h/r)_{max}$ occurs at $r=24r_g$ for $a_*=0$ and $r=7r_g$ for $a_*=1$].
If either $\epsilon_0 \dot M_o c^2$ or $\Delta \epsilon
\dot M_o c^2$ exceeds their respective limits, then the thin-disk approximation
breaks down.  In addition, if $h/r$ is small, then the approximation of a flat 
disk in the computation of the returning radiation will be appropriate.
To treat the interesting cases where $L \sim L_{Edd}$ will require a
2-D solution of the disk equations, which is beyond the scope of this
work.
The returning radiation will not affect the disk height since it 
diffuses through the disk on a thermal timescale, so there is no net flux 
due to returning radiation (unless the disk is warped).
\vskip 2mm
\hbox{~}
\centerline{\psfig{file=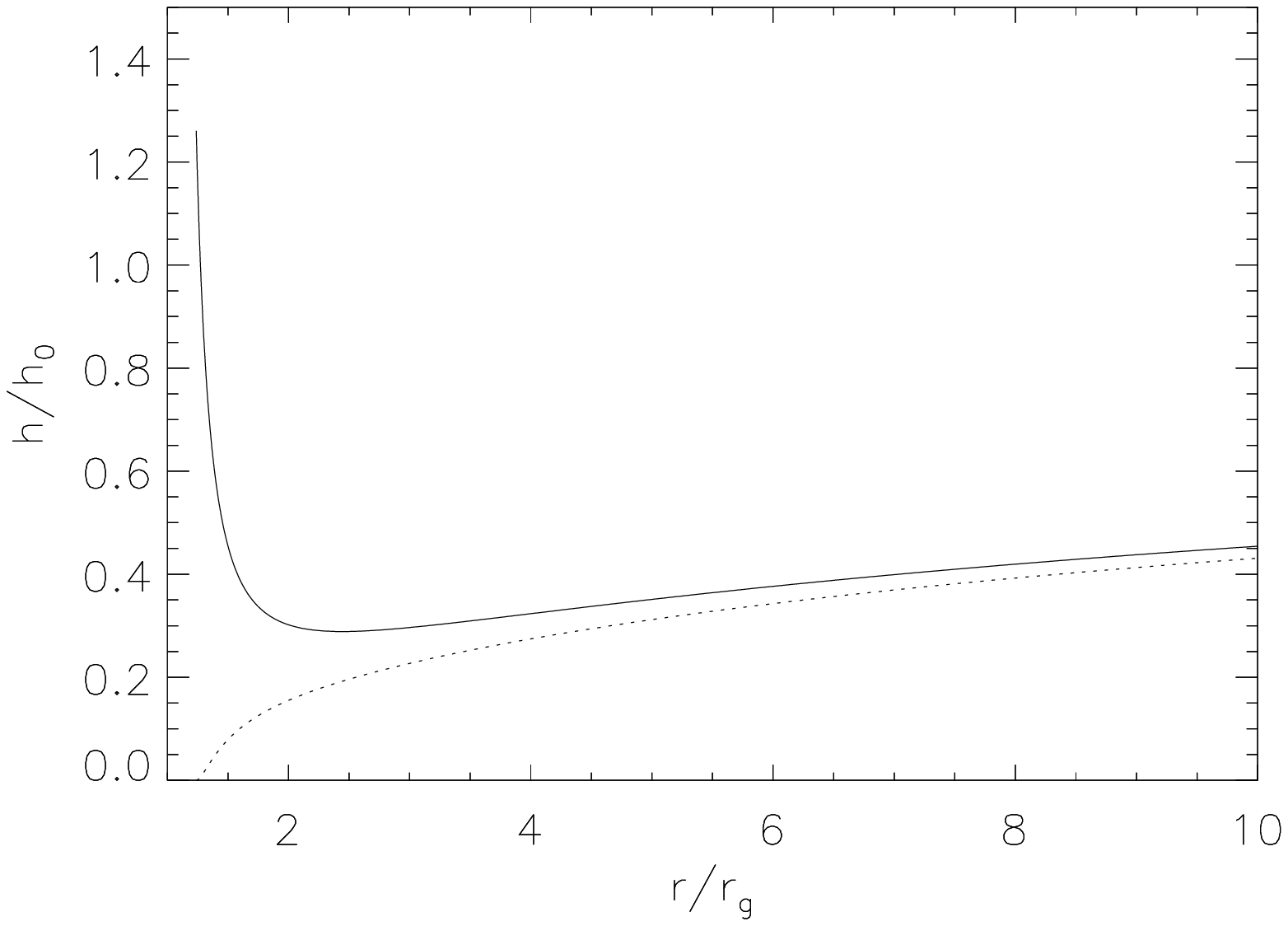,width=3.6in}} 
\noindent{
\scriptsize \addtolength{\baselineskip}{-3pt}
\vskip 1mm
\begin{normalsize}
Fig.~13.\  Height of radiation pressure-supported disk vs. radius
with $\epsilon = \epsilon_0$ (dotted line) and $\epsilon = 1$ (solid line)
for $a_*=0.998$.  The height is normalized by
$h_0 \equiv 3\kappa \dot M_o/(8\pi c)$.
\end{normalsize}
\vskip 3mm
\addtolength{\baselineskip}{3pt}
}
\vskip 2mm
\hbox{~}
\centerline{\psfig{file=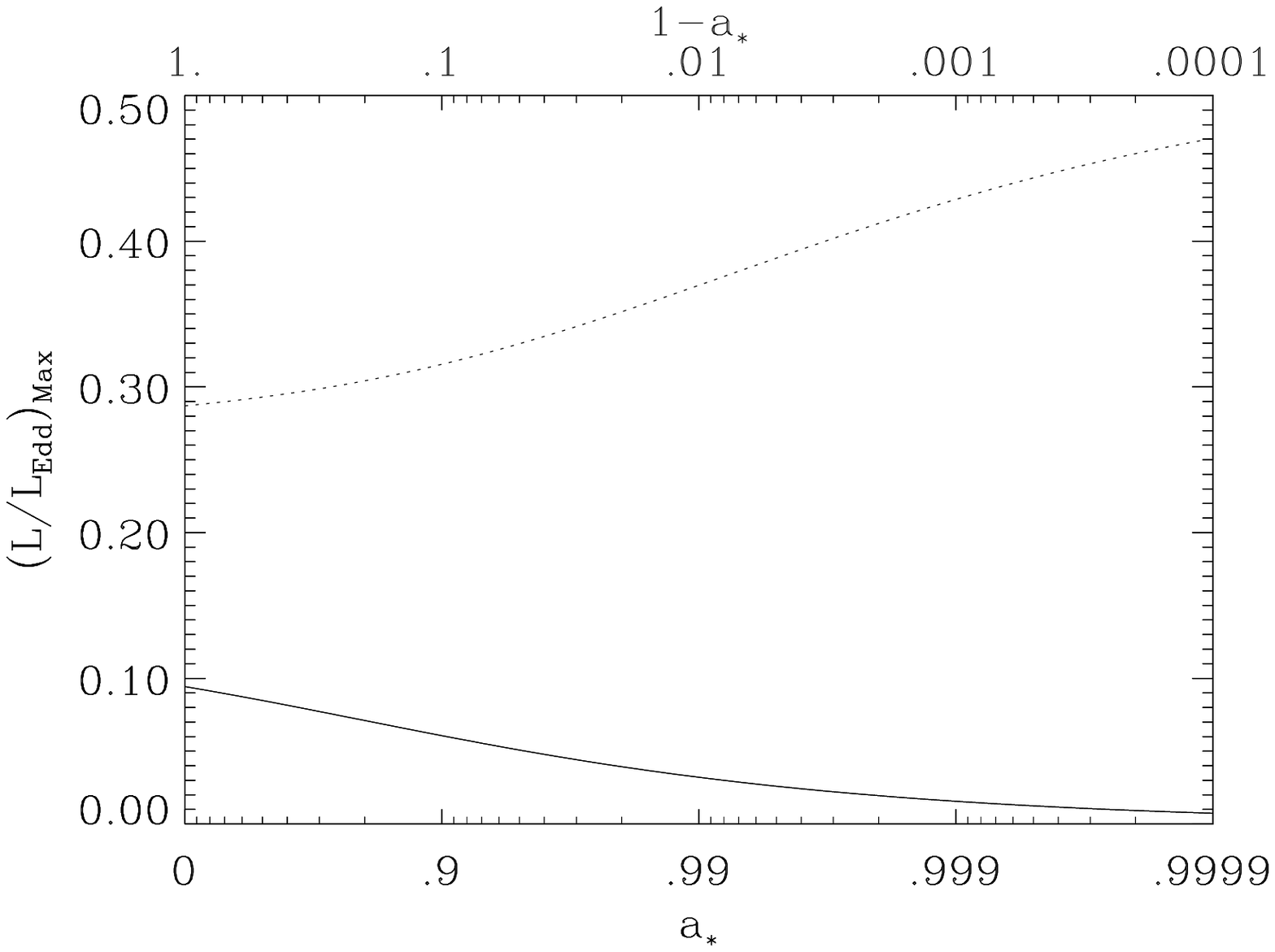,width=3.6in}} 
\noindent{
\scriptsize \addtolength{\baselineskip}{-3pt}
\vskip 1mm
\begin{normalsize}
Fig.~14.\  Upper limit on the luminosity for thin-disk ($h/r \lesssim 0.1$)
approximation to be valid at $r_{ms}$ for a radiation pressure-supported
disk.  Dotted line is for Novikov-Thorne disk; solid line is for
infinite-efficiency disk.
\end{normalsize}
\vskip 3mm
\addtolength{\baselineskip}{3pt}
}

\section{Conclusions}

        We have generalized the equations for an azimuthally symmetric,
geometrically thin, time-steady accretion disk around a black hole to
include the 
effects of a torque operating at the inner boundary, taken to be at 
$r_{ms}$.  Constant non-zero torque at $r_{ms}$ causes several
physical consequences that change the fundamental properties of the
accretion flow:

1)  The flux can be expressed as a sum of the usual Novikov \& Thorne
    expression plus a part due to the torque which scales roughly as $r^{-7/2}$.

2)  The accretion efficiency has a fundamental upper limit due to the 
    second law of black hole dynamics for $a_* < 0.36$.  For larger
    $a_*$, infinite efficiency is possible in principle.

3)  The black hole spin can reach an equilibrium for $a_* < 0.998$
    since the angular momentum reaching the hole is smaller.  Radiation 
    can also exert a significant torque on the black hole, which changes
    the value of the equilibrium-spin efficiency.  Above an 
    efficiency of $\epsilon = 0.36$, the black hole must always be spun down.

4)  Since the extra emissivity is peaked at the inner edge of the
    disk, if $r_{ms}$ is small then gravity causes a large fraction of 
    the radiation (up to 58\%) to return to the accretion disk.  The
    flux of returning radiation scales as $r^{-3}$ at large radius.  Up to
    15\% of the radiation can be captured by the black hole.

5)  The extra heating within the disk will increase the height of the
    disk if it is radiation pressure-supported.  This limits severely
    the luminosity at which the thin-disk approximation is appropriate.

6)  Doppler beaming and relativistic bending are strongest in the inner
    parts of the accretion disk where the extra flux peaks, so that
    for large $a_*$ and $\epsilon$, the disk will be limb-brightened.

These each have multiple observable consequences:

1)  The extra surface brightness changes the locally radiated spectrum.
Though the local surface brightness is usually not directly observable,
it may be possible to map it using several devices: eclipse mapping
(Baptista et al. 1998),
although no eclipsing black hole X-ray binaries have been discovered yet;
quasar microlensing (Agol \& Krolik 1999); or reverberation mapping
(Collier et al. 1999). 
If the Fe K$\alpha$ emissivity is proportional to the local dissipation,
these effects can strengthen the red wing of the line, particularly when
the spin is small.  This effect may undercut the argument that lines with
strong red wings came from disks around black holes with higher spin 
(e.g. Dabrowski et al. 1997).

Several authors have used the Novikov-Thorne model to fit
the soft X-ray spectra of galactic black hole candidates.  Their procedure
was to estimate the effective radiating area required to emit the observed
luminosity at the observed effective temperature.   On the basis of these 
fits they inferred that some black holes have rather high spins because the
effective radiating area of a Novikov-Thorne disk decreases with increasing 
spin (Zhang, Cui, \& Chen  1997).  However, for fixed spin and central mass, 
a disk with large $\Delta \epsilon$ has a smaller effective radiating area 
than a Novikov-Thorne disk, mimicking the effect of greater spin.

2)  The outer parts of accretion disks can be unstable to warping due to
irradiation from the center (Pringle 1996).  The minimum radius for growth 
of small warps is proportional to $\epsilon^{-2}$; if the efficiency is much 
higher than that of a standard disk, the minimum radius may be greatly shrunk.
Limb-brightening increases the effective efficiency and therefore makes
the linear growth more rapid; on the other hand, the corresponding
relative decrease in intensity away from the central disk plane may
weaken this effect in the non-linear regime.

3)  Wavelength-dependent limb-brightening (or limb-darkening)
introduces viewing angle-dependent biases into any flux-limited sample.
This is a particularly strong effect in the context of quasar surveys
because the number count distribution is so steep.  A variety of
distortions could occur in our view of what constitutes a ``typical"
quasar (cf. Krolik \& Voit 1998).

4)  Returning radiation can change the conditions for launching a
radiation-driven disk wind (e.g. Murray et al. 1995).  The returning
radiation is reradiated locally, so the net vertical force 
depends on the frequency-averaged opacities of the downgoing and upgoing 
radiation fields.  The radial component of the radiation force will
be larger than for a standard disk due to the higher efficiency
and limb brightening.

5)  Returning radiation causes the various annuli to ``communicate''
on the light crossing timescale.
Fluctuations of the flux at small radii will cause only 
slightly delayed fluctuations of the flux at larger radii, which emit 
at longer wavelengths.  Indeed, exactly this sort of behavior is commonly 
seen in accreting black hole systems.  For example, campaigns monitoring 
AGN have consistently found that continuum fluctuations are very nearly 
simultaneous all the way from $\simeq 1300$~\AA\  to $\simeq 5000$~\AA\ 
(Clavel et al. 1991; Korista et al. 1995; Wanders et al. 1997; Collier 
et al. 1998; O'Brien et al. 1998; Cutri et al. 1985).  Comparing the 
upper bounds on any inter-band delays to the radial scales expected on 
the basis of conventional disk models, these observations have been 
interpreted as requiring a coordinating signal group speed of at least 
$\sim 0.1c$ (e.g., Krolik et al. 1991; Courvoisier \& Clavel 1991;
Collier et al.  1998).

6)  The scattered component of returning radiation is highly
polarized parallel to the disk axis at high frequencies. This polarization 
rise may be related to the observed sharp rises in several
quasars (Koratkar et al. 1995, Impey et al. 1995). If the inner regions
of the disk have a strong Lyman continuum in emission, then this
scattered emission edge will appear as a strongly polarized, blueshifted
emission edge in the spectrum.  In an irradiated disk 
atmosphere there might be an additional effect: heating of the upper layers of 
the atmosphere can cause a temperature inversion, which changes the sense 
of the polarization.  We leave all such detailed calculations to future work.

	A question left unanswered by this work is what spin and 
efficiency we expect to be achieved by black holes in nature.  That there
is an upper limit on efficiency for an equilibrium spin 
means that a black hole with $\epsilon > \epsilon_{eq}$
must be born with original spin, or must be spun up
by accretion in which magnetic torques inside $r_{ms}$ do not play
an important role.  In the supermassive black hole case, the
spin may result from a merger.  No one has computed the final
spin of the resulting merger of two black holes; however, current
approximate calculations indicate that the final spin could be
quite large (Khanna et al. 1999).  The strength of the torque at
$r_{ms}$ depends on the strength of the magnetic field
in the accretion disk and the geometrical thickness of the flow,
which in turn depend on the accretion rate.  This dependence
will be best addressed with numerical simulations of MHD accretion.

\acknowledgments

   We would like to thank Roger Blandford, Doug Eardley, and Omer Blaes for
helpful conversations.  We would also like to thank the Institute for
Theoretical Physics at U.C., Santa Barbara for its hospitality.

This work was partially supported by NASA Grant NAG 5-3929 and NSF Grant
AST-9616922.

\appendix

\begin{deluxetable}{cccccccc}
\tablewidth{6.in}
\tablecaption{Coefficients for fits to $dM/dt$ and $dJ/dt$.}
\tablehead{\colhead{fit} & \colhead{$a_0$} & \colhead{$a_1$} & \colhead{$a_2$} & \colhead{$a_3$} &\colhead{$a_4$}& \colhead{$a_5$}& \colhead{$a_6$}}
\startdata
$10^4(dM/dt)_{rad}^0$ & 2.402&-11.63&10.87&-28.47&-17.30 &-3.651&-0.2731\nl
$10^4(dM/dt)_{rad}^\infty$  & 261.8&-600.9&449.7&554.9 &224.9&41.28&2.885\nl
$10^4(dJ/dt)_{rad}^0$ &.5182 &1.055&-5.149&3.707&2.347& .5455&.04527 \nl
$10^4(dJ/dt)_{rad}^\infty$  & 88.53& 29.56&-29.19&-200.6&-128.1&-30.00&-2.475\nl
\enddata \label{tab2}
\end{deluxetable}

The function $R_{ret}$ can be expressed as the sum of a part due
to the Novikov \& Thorne accretion rate, and a part due to the
torque at the inner edge:
\begin{equation}
R_{ret}(\epsilon,a_*,r) = R_0(a_*,r) + R_\infty(a_*,r) \Delta \epsilon.
\end{equation}
At large radius, these functions become constant, where we have fitted them
with polynomials in $x\equiv log_{10}(1-a_*)$:
\begin{eqnarray}
R_0(a_*,\infty)      &=& 0.0200 - 0.0360 x + 0.0279 x^2 + 0.00213 x^3 -0.00153 x^4 -0.000225 x^5, \cr
R_\infty(a_*,\infty) &=& 0.594 - 0.199  x - 0.116 x^2 - 0.107 x^3 - 0.0373 x^4 -0.00409 x^5.
\end{eqnarray}
The formulae are accurate to better 0.6\% from 
$a_*=0$ to $a_*=0.9999$.

For computing the spin evolution of an accreting black hole, it 
is useful to know $(dJ/dt)_{rad}=(dJ/dt)_{rad}^0 + \Delta
\epsilon (dJ/dt)_{rad}^\infty$ and $(dM/dt)_{rad}=(dM/dt)_{rad}^0 + \Delta
\epsilon (dM/dt)_{rad}^\infty$.   Note that $(dM/dt)^0_{rad} = M_1' + 
\epsilon_0 M_2'$ and $(dM/dt)^\infty_{rad} = M_2'$, and likewise for J.
We have fitted these 
with 6th order polynomials in $x$.  The coefficients $a_i$ of the
fits ($\sum_{i=0}^5 a_i x^i$)  are given in table \ref{tab2}, where we have 
multiplied each $a_i$ by $10^4$.

\end{document}